\documentclass[aps,pra,floatfix,showpacs]{revtex4}

\usepackage{graphicx}
\usepackage[english]{babel}
\usepackage{amsmath}
\usepackage{amssymb}
\usepackage{slashed}
\allowdisplaybreaks
\begin{document}
\title{Theories of Linear Response in BCS Superfluids and How They Meet Fundamental Constraints}

\author{Hao Guo$^{1,2}$, Chih-Chun Chien$^3$, Yan He$^{4,5}$}
\affiliation{$^1$Department of Physics, Southeast University, Nanjing 211189, China}
\affiliation{$^2$Department of Physics, University of Hong Kong, Hong Kong 999077, China}
\affiliation{$^3$Theoretical Division, Los Alamos National Laboratory, Los Alamos, NM, 87545, USA}
\affiliation{$^4$Department of Physics, University of California, Riverside, CA 92521, USA}
\affiliation{$^5$James Franck Institute, University of Chicago, Chicago, IL 60637, USA}
\begin{abstract} We address the importance of symmetry and symmetry breaking  on linear response theories of fermionic
BCS superfluids. The linear theory of a noninteracting Fermi gas is reviewed and several consistency constraints are verified.
The challenge to formulate linear response theories of BCS superfluids consistent with density and spin conservation laws comes from
 the presence of a broken
U(1)$_{\textrm{EM}}$ symmetry associated with electromagnetism (EM) and we discuss two routes for circumventing this.
The first route follows Nambu's integral-equation approach for the EM vertex function, but this method is not specific for BCS superfluids.
We focus on the second route based on a consistent-fluctuation-of-the order-parameter (CFOP)
approach where the gauge transformation and the fluctuations of the order parameter
are treated on equal footing.
The CFOP approach allows one to explicitly verify several important constraints:
The EM vertex satisfies not only a  Ward identity which guarantees charge conservation
but also a $Q$-limit Ward identity associated with
the compressibility sum rule.
In contrast, the spin degrees of freedom associated with another
U(1)$_z$ symmetry are not affected by the Cooper-pair condensation that breaks only the U(1)$_{\textrm{EM}}$ symmetry.
As a consequence the collective modes from the fluctuations of the order parameter only couple to the density response function but decouple from the spin response function, which reflects the different fates of the two U(1) symmetries in the superfluid phase.
Our formulation lays the ground work for application to more general
theories of BCS-Bose Einstein
Condensation (BEC) crossover both above and below $T_c$.
\end{abstract}


\maketitle

\section{Introduction}
Linear response theories have been an important tool for studying transport and dynamic properties of many-body systems \cite{Kubo_book,Walecka,Mahanbook}.
Although there have been myriad successful examples in classical or non-interacting systems, developing a linear response theory for
complex systems or strongly correlated systems could be quite a challenge. As summarized in Ref.~\cite{Mahanbook}, it is difficult to obtain consistent expressions of the compressibility of an interacting electron gas from the derivatives of thermodynamic quantities and from the correlation functions.
A naive calculation of the response functions of BCS theory of conventional superconductors could have led to a violation of the charge conservation. The spontaneous symmetry breaking of the superfluid phase further complicates the treatment of any linear response theory of BCS superfluids.
Thus it requires more sophisticated treatments \cite{Nambu60,Kadanoff61} to obtain the response functions that respect conservation laws.

Here we analyze two consistent linear response theories of BCS superfluids in great details and review the importance of gauge invariance, the (generalized) Ward
identity, sum rules, and how they impose constraints on response functions.
Also crucial is an additional
$Q$-limit Ward identity \cite{Maebashi09}
which is related to the compressibility sum rule.
The non-interacting Fermi gas satisfies all constraints, albeit in a trivial sense. In the presence of interactions,
we will focus on a linear response theory which we call
consistent-fluctuation-of-the order-parameter
(CFOP) theory for fermionic superfluids. Importantly, at the mean-field level,
this approach is consistent with all conservation laws and associated sum rules. This will set up a solid foundation for the generalization to the more general theories of BCS-BEC crossover \cite{OurAnnPhys}.
We also present a spin linear
response theory in a systematic manner to demonstrate that fundamental constraints including the sum rules and Ward identities are also satisfied in the spin channel.
One major difference between the density and spin response functions
is that in the superfluid phase the order parameter, which corresponds to the condensate of Cooper pairs, breaks only the symmetry associated with the density response functions, but not
the symmetry associated with the spin response function. Therefore the density response function exhibits richer structures across the superfluid
phase transition. This is the base for constructing an order-parameter like quantity that measures the ``spin-charge separation" in a BCS superfluid as proposed in Ref.~\cite{HaoPRL10}.

There have been two major approaches for addressing the linear response
of fermionic BCS superfluids.
The first one is attributed to Nambu \cite{Nambu60}, who
reformulated the problem in the two-dimensional Nambu space and pointed out that if
certain types of corrections to the EM interacting vertex are considered consistently with those corrections to the self-energy in the Green's function, the gauge invariance can be maintained explicitly. Nambu presented a generalized form of Ward
identity (GWI) and proposed an integral equation for the full electromagnetic (EM) interacting vertex
 in the Nambu space. He stated that this EM vertex must satisfy the GWI without giving a proof
 and we present our own proof in Appendix~\ref{app:b0}. By solving this integral equation at small
frequency and momentum limits, it was shown that the excitations of collective modes correspond to the poles of the density response function.
Hence the many-particle effects are indeed included in the corrections of EM vertex.
Despite many virtues, this method is relatively difficult to implement. Firstly, it is very
hard to solve the integral equation for the vertex. If one truncates the integral equation, one may not get a gauge invariant solution and can not even find the correct
collective-mode excitations. Secondly, this approach was formulated in the Nambu space and it is difficult to translate the results in the
two-dimensional Nambu space to their counterparts in the one-dimensional representation space of
fermion operators. This limits its applicability to more general problems such as BCS-BEC crossover.
Moreover, it can be shown that the GWI discussed in Nambu's original paper \cite{Nambu60} is not a unique constraint for gauge invariance of BCS theory. In other words, Nambu's method is not specific to BCS
theory: It is more general but harder to obtain an exact solution.

There is yet another approach which we call the CFOP approach with a totally different structure
when compared to Nambu's method. Our goal is to explain this approach and test its results against
some fundamental constraints. Within this approach, we treat the gauge transformation and the
fluctuations of order parameter on equal footing, such that many-particle effects are
also explicitly included in the EM vertex. This approach was first proposed by Kadanoff and Martin\cite{Kadanoff61} in a less
complete form. They only considered the fluctuations of the order parameter, and
tried to decompose the three-particle Green's functions in a way that can respect
gauge invariance. It was then independently formulated in several unrelated papers by
Betbeder Matibet and Nozieres \cite{Nozieres69} and Kulik et al. \cite{KulikJLTP81}
in a more complete form where both phase and amplitude fluctuations are considered.
Later on, Zha and Levin\cite{ZhaPRB95} revisited it and presented three identities of
response functions, which are now known to be part of the WIs for response functions.
Recently, it was again proposed in the Keldysh formalism with time-ordered Green's
functions in Ref. \cite{Arseev}. A similar derivation using a kinetic-equation formalism
is also discussed in Ref.~\cite{Stringari06}.
 In this paper, we cast this formalism in a more
systematic and covariant form. We will present more virtues of this approach, such as
the consistency with the $Q$-limit Ward identity and the $f$-sum rule. The EM vertex and
its generalized form can be both found in different representation spaces.
A comparison between
Nambu's method and the CFOP theory will also be presented. We will show that the CFOP approach
is indeed a consistent and manageable linear response theory for BCS superfluids.
The generalization of the
CFOP theory to relativistic BCS superfluids can be found in Ref.~\cite{OurPRD12}.
Besides linear response theories discussed here, one may find more discussions on the gauge invariance of BCS superconductors in terms of effective field theories in Ref.~\cite{Greiter}.

This paper is organized as follows.
In Section~\ref{C1} we briefly discuss the general symmetries in the theory
of two-component Fermi gases with contact interactions.
In Section~\ref{C2} we review the
density and spin linear response theories for non-interacting Fermi gases. In Section.\ref{C4} we spend a significant part of the paper on
the CFOP linear response theory of BCS superfluids. More specifically,
Section~\ref{CS2} presents the derivation and some results of the CFOP linear
response theory; Section~\ref{CS3} addresses the (generalized) Ward
identity and $Q$-limit (generalized) Ward identity of the CFOP linear response theory; Sections~\ref{4C} and \ref{CS6} review Nambu's integral-equation approach and present
a comparison of our CFOP approach with Nambu's method.
In Section~\ref{C5} we develop a parallel formalism for the linear
response theory in the spin channel by ``gauging'' the U(1)$_z$ symmetry. Section~\ref{Conclusion} concludes our work.

\section{Symmetries of Theories of Fermi Gases with Contact Interactions}\label{C1}
Throughout this paper, we follow the convention $e=c=\hbar=1$ and use $\sigma$ to denote the spin or pseudo-spin $\uparrow,\downarrow$ with
$\uparrow$ and $\bar{\sigma}$ being the opposite of $\downarrow$ and $\sigma$ respectively. The metric tensor of the Minkowski spacetime is chosen as $\eta^{\mu\nu}=\textrm{diag}(1,-1,-1,-1)$
and Einstein summation convention is adopted.
For a two component Fermi gas interacting via contact interactions, we consider the Hamiltonian
\begin{equation}\label{H00}
H=\int d^3\mathbf{x}\psi^{\dagger}_{\sigma}(\mathbf{x})\Big(\frac{\hat{\mathbf{p}}^2}{2m}-\mu\Big)\psi_{\sigma}(\mathbf{x})-g\int
d^3\mathbf{\mathbf{x}}\psi^{\dagger}_{\uparrow}(\mathbf{x})\psi^{\dagger}_{\downarrow}(\mathbf{x})\psi_{\downarrow}(\mathbf{x})\psi_{\uparrow}(\mathbf{x}),
\end{equation}
where $\psi$ and $\psi^{\dagger}$ are the annihilation and creation operators of fermions, $\mu$ is the chemical potential, $m$ is the fermion mass,
and $g$ is the bare coupling constant. There is an implicit summation over the pseudo-spin indices $\sigma$. The Hamiltonian (\ref{H00}) has a
SU(2)$\times$SU(2) symmetry \cite{Nambu60,CNYangPRL89,CNYangMPLB90}. The first SU(2) symmetry is generated by
\begin{eqnarray}\label{T1}
& &\psi_{\uparrow}\rightarrow \textrm{cosh}\chi\psi_{\uparrow}+\textrm{sinh}\chi\psi^{\dagger}_{\downarrow},\quad
\psi^{\dagger}_{\downarrow}\rightarrow \textrm{sinh}\chi\psi^{\dagger}_{\downarrow}+\textrm{cosh}\chi\psi_{\uparrow}, \label{u1}\nonumber\\
& &\psi_{\uparrow}\rightarrow \textrm{cosh}\chi\psi_{\uparrow}-i\textrm{sinh}\chi\psi^{\dagger}_{\downarrow},\quad
\psi^{\dagger}_{\downarrow}\rightarrow i\textrm{sinh}\chi\psi^{\dagger}_{\downarrow}+\textrm{cosh}\chi\psi_{\uparrow}, \label{u2}\nonumber\\
& &\psi_{\sigma}\rightarrow e^{-i\chi}\psi_{\sigma},\quad \psi^{\dagger}_{\sigma}\rightarrow e^{i\chi}\psi^{\dagger}_{\sigma}, \label{u3}
\end{eqnarray}
where $\chi$ is a continuous parameter. The transformation on the third line is the well-known U(1) symmetry associated to electromagnetism if the
system is charged. The generators of the these transformations are $-i\sigma_1$, $-i\sigma_2$ and $\sigma_3$ in the space spanned by the Nambu spinor $(\psi_{\uparrow},\psi^{\dagger}_{\downarrow})^T$, where $\sigma_i$ are the Pauli matrices. Hence the symmetry (\ref{T1}) is indeed SU(2), or more
precisely, SU(1,1) with the U(1)$_{\textrm{EM}}$ being its subgroup.

The second SU(2) symmetry is given by
\begin{eqnarray}\label{T2}
& &\psi_{\uparrow}\rightarrow \textrm{cos}\alpha\psi_{\uparrow}+i\textrm{sin}\alpha\psi_{\downarrow},\quad \psi^{\dagger}_{\downarrow}\rightarrow
-i\textrm{sin}\alpha\psi^{\dagger}_{\uparrow}+\textrm{cos}\alpha\psi^{\dagger}_{\downarrow}, \label{sx}\nonumber\\
& &\psi_{\uparrow}\rightarrow \textrm{cos}\alpha\psi_{\uparrow}+\textrm{sin}\alpha\psi_{\downarrow},\quad \psi^{\dagger}_{\downarrow}\rightarrow
-\textrm{sin}\alpha\psi^{\dagger}_{\uparrow}+\textrm{cos}\alpha\psi^{\dagger}_{\downarrow}, \label{sy}\nonumber\\
& &\psi_{\sigma}\rightarrow e^{-iS_{\sigma}\alpha}\psi_{\sigma},\quad \psi^{\dagger}_{\sigma}\rightarrow e^{iS_{\sigma}\alpha}\psi^{\dagger}_{\sigma},
\textrm{ where } S_{\uparrow,\downarrow}=\pm 1. \label{sz}
\end{eqnarray}
The generators are $\sigma_1$, $\sigma_2$ and $\sigma_3$ in the space spanned by $(\psi_{\uparrow},\psi_{\downarrow})^T$. The transformation on the
third line is the spin rotation around the $z-$axis, which forms the subgroup U(1)$_{z}$ of the second SU(2). In what follows, we will focus on the
U(1)$_{\textrm{EM}}\times$U(1)$_{z}$ symmetries, and we may call the theories associated with them as being in the density and spin channels respectively.

When the continuous parameter $\chi$ become space-time dependent, the two global U(1) symmetries are ``gauged'' respectively. To keep the Lagrangian invariant under the gauge transformations, one needs to couple the fermion field by an external vector field $A^{\mu}=(\phi,\mathbf{A})$ which
transforms as $\mathbf{A}\rightarrow \mathbf{A}-\nabla\chi$, $\phi\rightarrow \phi+\frac{\partial \chi}{\partial t}$. For the U(1)$_{\textrm{EM}}$
symmetry, the Hamiltonian (\ref{HBCS}) becomes
\begin{eqnarray}\label{EMH}
& &H=\int
d^3\mathbf{x}\psi^{\dagger}_{\sigma}(\mathbf{x})\Big(\frac{(\hat{\mathbf{p}}-\mathbf{A}(\mathbf{x}))^2}{2m}-\mu\Big)\psi_{\sigma}(\mathbf{x})+\int
d^3\mathbf{x}\phi(\mathbf{x})\psi^{\dagger}_{\sigma}(\mathbf{x})\psi_{\sigma}(\mathbf{x})-g\int
d^3\mathbf{\mathbf{x}}\psi^{\dagger}_{\uparrow}(\mathbf{x})\psi^{\dagger}_{\downarrow}(\mathbf{x})\psi_{\downarrow}(\mathbf{x})\psi_{\uparrow}(\mathbf{x})
\nonumber\\
&=&\int d^3\mathbf{x}\psi^{\dagger}_{\sigma}(\mathbf{x})\Big(\frac{\hat{\mathbf{p}}^2}{2m}-\mu\Big)\psi_{\sigma}(\mathbf{x})+\int
d^3\mathbf{x}\mathbf{J}(\mathbf{x})\cdot \mathbf{A}(\mathbf{x})+\int d^3\mathbf{x}n(\mathbf{x})\phi(\mathbf{x})-g\int
d^3\mathbf{\mathbf{x}}\psi^{\dagger}_{\uparrow}(\mathbf{x})\psi^{\dagger}_{\downarrow}(\mathbf{x})\psi_{\downarrow}(\mathbf{x})\psi_{\uparrow}(\mathbf{x}),
\end{eqnarray}
where
\begin{eqnarray}\label{EC}
&
&\mathbf{J}(\mathbf{x})=-\frac{1}{2mi}\Big[\psi^{\dagger}_{\sigma}(\mathbf{x})\big(\nabla\psi_{\sigma}(\mathbf{x})\big)-\big(\nabla\psi^{\dagger}_{\sigma}(\mathbf{x})\big)\psi_{\sigma}(\mathbf{x})\Big]-\frac{1}{m}\mathbf{A}(\mathbf{x})\psi^{\dagger}_{\sigma}(\mathbf{x})\psi_{\sigma}(\mathbf{x}),
\nonumber\\
& &n(\mathbf{x})=\psi^{\dagger}_{\sigma}(\mathbf{x})\psi_{\sigma}(\mathbf{x}).
\end{eqnarray}
Here $n(\mathbf{x})$ is the particle number density and $\mathbf{J}(\mathbf{x})$ is the mass current. There is also an implicit summation over the
repeated Greek index $\sigma$. When the external field $A^{\mu}$ is the EM field, the Noether current for the global U(1)$_{\textrm{EM}}$
symmetry is $J^{\mu}=(n, \mathbf{J})$ which obeys the conservation law $\partial_{\mu}J^{\mu}=0$ in the Heisenberg picture, where
$\partial_{\mu}=(\frac{\partial}{\partial t},\nabla)$.

The same discussion can be implemented in the spin channel. The fermion can also couple to an effective external
field $A^{\mu}$. Hence the Hamiltonian with a U(1)$_z$ ``gauge'' symmetry is given by
\begin{eqnarray}\label{spinH}
& &H=\int
d^3\mathbf{x}\psi^{\dagger}_{\sigma}(\mathbf{x})\Big(\frac{(\hat{\mathbf{p}}-S_{\sigma}\mathbf{A}(\mathbf{x}))^2}{2m}-\mu\Big)\psi_{\sigma}(\mathbf{x})+\int
d^3\mathbf{x}\phi(\mathbf{x})S_{\sigma}\psi^{\dagger}_{\sigma}(\mathbf{x})\psi_{\sigma}(\mathbf{x})-g\int
d^3\mathbf{\mathbf{x}}\psi^{\dagger}_{\uparrow}(\mathbf{x})\psi^{\dagger}_{\downarrow}(\mathbf{x})\psi_{\downarrow}(\mathbf{x})\psi_{\uparrow}(\mathbf{x})
\nonumber\\
&=&\int d^3\mathbf{x}\psi^{\dagger}_{\sigma}(\mathbf{x})\Big(\frac{\hat{\mathbf{p}}^2}{2m}-\mu\Big)\psi_{\sigma}(\mathbf{x})+\int
d^3\mathbf{x}\mathbf{J}_{\textrm{S}}(\mathbf{x})\cdot \mathbf{A}(\mathbf{x})+\int d^3\mathbf{x}n_{\textrm{S}}(\mathbf{x})\phi(\mathbf{x})-g\int
d^3\mathbf{\mathbf{x}}\psi^{\dagger}_{\uparrow}(\mathbf{x})\psi^{\dagger}_{\downarrow}(\mathbf{x})\psi_{\downarrow}(\mathbf{x})\psi_{\uparrow}(\mathbf{x}),
\end{eqnarray}
where
\begin{eqnarray}\label{SC}
&
&\mathbf{J}_{\textrm{S}}(\mathbf{x})=-\frac{1}{2mi}S_{\sigma}\Big[\psi^{\dagger}_{\sigma}(\mathbf{x})\big(\nabla\psi_{\sigma}(\mathbf{x})\big)-\big(\nabla\psi^{\dagger}_{\sigma}(\mathbf{x})\big)\psi_{\sigma}(\mathbf{x})\Big]-\frac{1}{m}\mathbf{A}(\mathbf{x})\psi^{\dagger}_{\sigma}(\mathbf{x})\psi_{\sigma}(\mathbf{x}),
\nonumber\\
& &n_{\textrm{S}}(\mathbf{x})=S_{\sigma}\psi^{\dagger}_{\sigma}(\mathbf{x})\psi_{\sigma}(\mathbf{x}).
\end{eqnarray}
Here $S_{\uparrow,\downarrow}=\pm1$ as in (\ref{sz}). The external
field $A_{\mu}$ has a different physical meaning from that in the density
channel. Since $n_{\textrm{S}}$ is the $z$ component of spin density, the field
$\phi$ coupled to $n_{\textrm{S}}$ corresponds to $B_z$.
$\mathbf{J}_{\textrm{S}}$ is the difference between the spin-up and the spin-down
currents, i.e., the magnetization current. Therefore the field which couples to it is
the magnetizing field $\mathbf{A}\equiv\mathbf{m}$. The effective external vector
field is thus $A^{\mu}\equiv(B_z,\mathbf{m})$. The resulting Hamiltonian then describes a generalized
spin-magnetic field interaction. The Noether current for the global U(1)$_z$ symmetry is
$J_{\textrm{S}}^{\mu}=(n_{\textrm{S}},\mathbf{J}_{\textrm{S}})$, which also satisfies
the conservation law $\partial_{\mu}J_{\textrm{S}}^{\mu}=0$.

\section{Linear Response Theory for Non-interacting Fermi Gases}
\label{C2}
We begin with the non-interacting Fermi gases where $g=0$. Many important identities can
be verified explicitly and they provide useful hints for the development of consistent theories of linear response for interacting Fermi gases.
When a non-interacting two-component Fermi gas is perturbed by a weak external gauge field, one can discuss the
response in the density channel as well as in the spin channel the leading-order or linear-approximation of the perturbation.

\begin{figure}
\begin{center}
\includegraphics[width=2.in,clip]
{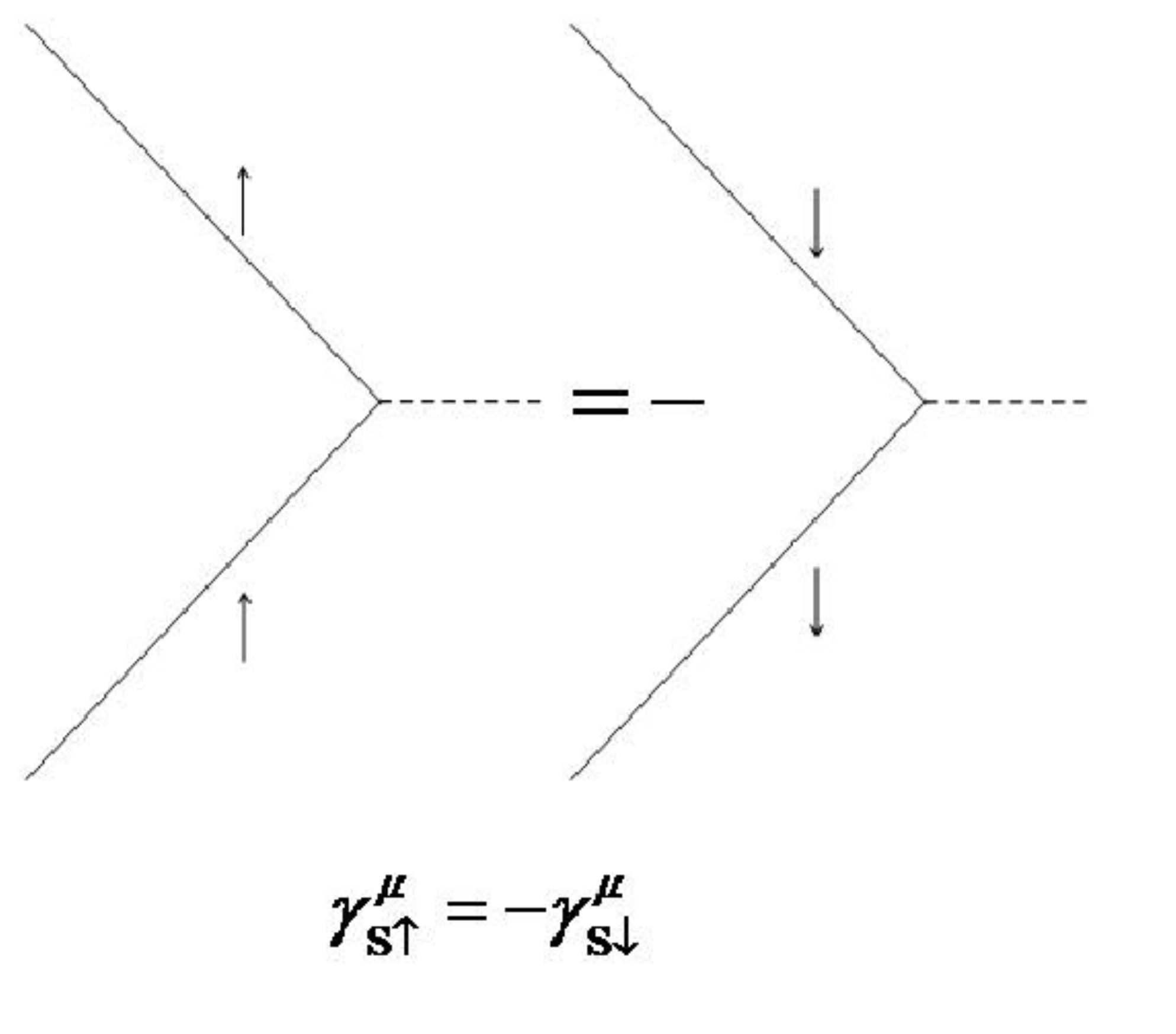}
\caption{The diagrams for the spin vertex function. Different spin indices have difference signs.}
\label{fig:S-V}
\end{center}
\end{figure}

In momentum space, the linear response theories associated with the two U(1) symmetries mentioned above
 can be expressed in a unified form as $H=H_0+H'_{\textrm{D,S}}$, where
\begin{eqnarray}
& &H_0=\sum_{\mathbf{p}}\psi^{\dagger}_{\sigma\mathbf{p}}\xi_{\mathbf{p}}\psi_{\sigma\mathbf{p}},\nonumber\\
& &H'_{\textrm{D}}=\sum_{\mathbf{p}\mathbf{q}}\psi^{\dagger}_{\sigma\mathbf{p}+\mathbf{q}}\gamma^{\mu}(\mathbf{p}+\mathbf{q},\mathbf{p})A_{\mu\mathbf{q}}\psi_{\sigma\mathbf{p}},\quad
H'_{\textrm{S}}=\sum_{\mathbf{p}\mathbf{q}}\psi^{\dagger}_{\sigma\mathbf{p}+\mathbf{q}}\gamma^{\mu}_{\textrm{S}\sigma}(\mathbf{p}+\mathbf{q},\mathbf{p})A_{\mu\mathbf{q}}\psi_{\sigma\mathbf{p}}.
\end{eqnarray}
Here $\xi_{\mathbf{p}}=\frac{p^2}{2m}-\mu$,
 and we have introduced the interacting vertices
$\gamma^{\mu}(\mathbf{p}+\mathbf{q},\mathbf{p})
=(1,\frac{\mathbf{p}+\frac{\mathbf{q}}{2}}{m})$
for the density channel and
$\gamma^{\mu}_{\textrm{S}\sigma}(\mathbf{p}+\mathbf{q},\mathbf{p})=(S_{\sigma},S_{\sigma}\frac{\mathbf{p}+\frac{\mathbf{q}}{2}}{m})$
for the spin channel. The latter explicitly depends on the spin (or pseudo-spin) and
Fig.\ref{fig:S-V} illustrates this spin dependence.  The
four-currents of the density and spin channels in momentum space are given by \begin{eqnarray}\label{Jds}
J^{\mu}_{\mathbf{q}}=\sum_{\mathbf{p}}\psi^{\dagger}_{\sigma\mathbf{p}}\gamma^{\mu}(\mathbf{p}+\mathbf{q},\mathbf{p})\psi_{\sigma\mathbf{p}+\mathbf{q}},\quad
J^{\mu}_{\textrm{S}\mathbf{q}}=\sum_{\mathbf{p}}\psi^{\dagger}_{\sigma\mathbf{p}}\gamma^{\mu}_{\textrm{S}\sigma}(\mathbf{p}+\mathbf{q},\mathbf{p})\psi_{\sigma\mathbf{p}+\mathbf{q}}.
\end{eqnarray}
Note in the perturbed Hamiltonian $H'_{\textrm{D,S}}$, we only keep the terms up to the linear order of the external field $A^{\mu}$ and higher order terms are neglected.
Hence in the expressions of the currents $J^{\mu}_{\mathbf{q}}$ and $J^{\mu}_{\textrm{S}\mathbf{q}}$ the linear terms of $A^{\mu}$ are dropped,
which is different from the currents given by Eqs.~(\ref{EC}) and (\ref{SC}).

For the rest of this section we will focus on the density channel since
the linear response theory in the spin channel has a similar structure as the former, and the response kernels can be shown to be the same.
The imaginary time $\tau=it$ is introduced here and the Heisenberg operator is
defined as $\mathcal{O}(\tau)=e^{H\tau}\mathcal{O}e^{-H\tau}$. The spacetime
coordinates are $x=(\tau,\mathbf{x})$. Hence the Green's function is defined by
$G_0(x,x')=-\langle T_{\tau}[\psi_{\sigma}(x)\psi_{\sigma}^{\dagger}(x')]\rangle$
where $T_{\tau}$ denotes the $\tau$-order of operators. Its expression in momentum
space is $G_0(i\omega_n,\mathbf{p})=(i\omega_n-\xi_{\mathbf{p}})^{-1}$. The four-momentum
is defined as $P\equiv p_{\mu}=(i\omega_n,\mathbf{p})$ where $\omega_n=(2n+1)\pi
k_BT$ is the fermion Matsubara frequency. The particle density is given by
\begin{eqnarray}\label{nef}
n=T\sum_{i\omega_n}\sum_{\mathbf{p},\sigma}G_0(P)=2\sum_{\mathbf{p}}f(\xi_{\mathbf{p}}),
\end{eqnarray} where $f(x)=1/(e^{x/k_B T}+1)$ is the fermion distribution function. By defining $h^{\mu\nu}=-\eta^{\mu\nu}(1-\eta^{\nu0})$, we formally write the perturbation of the EM current
$J^{\mu}$ as \begin{eqnarray}
\delta J^{\mu}(\tau,\mathbf{q})=\sum_{\mathbf{p}}\langle\psi^{\dagger}_{\sigma\mathbf{p}}(\tau)\gamma^{\mu}(\mathbf{p}+\mathbf{q},\mathbf{p})\psi_{\sigma\mathbf{p}+\mathbf{q}}(\tau)\rangle+\frac{n}{m}h^{\mu\nu}A_{\nu}(\tau,\mathbf{q}),
\end{eqnarray}
The extra term
 $\frac{n}{m}h^{\mu\nu}$ of $\delta J^{\mu}$ arises from the second term of $\mathbf{J}$ or $\mathbf{J}_{\textrm{S}}$ (See Eqs.(\ref{EC}) or (\ref{SC})), since
$\mathbf{J}$ or $\mathbf{J}_{\textrm{S}}$ already contains a first order term of $A^{\mu}$. The linear response
theory is then written in the form
\begin{eqnarray}
\delta J^{\mu}(\tau,\mathbf{q})=\int d\tau'\big(Q^{\mu\nu}_0(\tau-\tau',\mathbf{q})+\frac{n}{m}h^{\mu\nu}\delta(\tau-\tau')\big)A_{\nu}(\tau',\mathbf{q}),
\end{eqnarray}
where the response kernels are \begin{eqnarray}\label{Qmn}
Q^{\mu\nu}_0(\tau-\tau',\mathbf{q})&=&-\langle
T_{\tau}[J^{\mu}(\tau,\mathbf{q})J^{\nu}(\tau',-\mathbf{q})]\rangle\nonumber\\
&=&-\sum_{\mathbf{p}\mathbf{p}'}\langle
T_{\tau}[\psi^{\dagger}_{\sigma\mathbf{p}}(\tau)\gamma^{\mu}(\mathbf{p}+\mathbf{q},\mathbf{p})\psi_{\sigma\mathbf{p}+\mathbf{q}}(\tau)\psi^{\dagger}_{\sigma'\mathbf{p}'+\mathbf{q}}(\tau')\gamma^{\nu}(\mathbf{p}',\mathbf{p}'+\mathbf{q})\psi_{\sigma'\mathbf{p}'}(\tau')]\rangle.
\end{eqnarray}
Implementing a Fourier transform and making use of Wick's theorem, we obtain
 \begin{eqnarray}\label{Qmn0}
Q^{\mu\nu}_0(i\Omega_l,\mathbf{q})&=&2T\sum_{i\omega_n}\sum_{\mathbf{p}\mathbf{p}'}\gamma^{\mu}(\mathbf{p}+\mathbf{q},\mathbf{p})G_{0\mathbf{p}+\mathbf{q},\mathbf{p}'+\mathbf{q}}(i\omega_n+i\Omega_{l})
\gamma^{\nu}(\mathbf{p}',\mathbf{p}'+\mathbf{q})G_{0\mathbf{p},\mathbf{p}'}(i\omega_n)\nonumber\\
&=&2T\sum_{i\omega_n}\sum_{\mathbf{p}}\gamma^{\mu}(P+Q,P)G_{0}(P+Q)
\gamma^{\nu}(P,P+Q)G_{0}(P). \end{eqnarray} where $Q\equiv
q_{\mu}=(i\Omega_{l},\mathbf{q})$, $\Omega_{l}$ is the boson Matsubara frequency, and
$G_{0\mathbf{p},\mathbf{p}'}(i\omega_n)=G_{0\mathbf{p}}(\omega_n)\delta_{\mathbf{p},\mathbf{p}'}$
with $G_{0\mathbf{p}}(\omega_n)\equiv G_0(P)$. For convenience, we have defined
$\gamma^{\mu}(P+Q,P)\equiv\gamma^{\mu}(\mathbf{p}+\mathbf{q},\mathbf{p})$. The factor
2 comes from the summation over the spin (or pseudo-spin) indices $\sigma$. The spin response
kernels have the same structure as one can see from the facts that
$\gamma^{\mu}_{\textrm{S}\sigma}(\mathbf{p}+\mathbf{q},\mathbf{p})=S_{\sigma}\gamma^{\mu}(\mathbf{p}+\mathbf{q},\mathbf{p})$
and $S^2_{\sigma}=1$.
Note there are also two spin interacting vertices inside
the expression of the spin response kernels similar to (\ref{Qmn}).
The detailed expressions of the response functions
are listed in Appendix~\ref{app:a0}, where
$\xi^{\pm}_{\mathbf{p}}=\xi_{\mathbf{p}\pm\frac{\mathbf{q}}{2}}$.

Due to the simple structure, one may verify the following identities explicitly: (1) Ward identities (WIs), (2) compressibility sum rule, (3) $f$-sum rule, and (4) $Q$-limit Ward identity.
By direct evaluating each term, one can verify that the EM and spin interacting vertices satisfy the WIs
\begin{eqnarray}\label{BWI}
q_{\mu}\gamma^{\mu}(P+Q,P)&=&G^{-1}_0(P+Q)-G^{-1}_0(P),\nonumber\\
q_{\mu}\gamma^{\mu}_{\textrm{S}\sigma}(P+Q,P)&=&S_{\sigma}\big(G^{-1}_{0}(P+Q)-G^{-1}_{0}(P)\big).
\end{eqnarray} This leads to the WIs for the response kernels
$q_{\mu}\tilde{Q}^{\mu\nu}_0(Q)=0$ where
$\tilde{Q}^{\mu\nu}_0=Q^{\mu\nu}_0+\frac{n}{m}h^{\mu\nu}$. It can be shown as follows
\begin{eqnarray}\label{WIP}
q_{\mu}\tilde{Q}^{\mu\nu}_0(Q)
&=&2\sum_{P}[G^{-1}_0(P+Q)-G^{-1}_0(P)]G_0(P+Q)\gamma^{\nu}(P,P+Q)G_0(P)-\frac{n}{m}q^{\nu}(1-\eta^{\nu0})\nonumber\\
&=&2\sum_{P}G_0(P)[\gamma^{\nu}(P+Q,P)-\gamma^{\nu}(P-Q,P)]-\frac{n}{m}q^{\nu}(1-\eta^{\nu0})\nonumber\\
&=&2\sum_{P}G_0(P)\frac{q^{\nu}}{m}(1-\eta^{\nu0})-\frac{n}{m}q^{\nu}(1-\eta^{\nu0})=0,
 \end{eqnarray} where $\sum_P\equiv T\sum_{i\omega_n}\sum_{\mathbf{p}}$. The WIs for the response kernels further lead to
the conservation of the perturbed current $q_{\mu}\delta J^{\mu}(Q)=0$. For the spin
response, the same conclusions can be obtained.

Next we show that the response function satisfies the compressibility sum rule
\cite{Kubo_book} and $f$-sum rule
\begin{eqnarray}\label{SRF} & &\frac{\partial
n}{\partial \mu}=-Q^{00}_0(\omega=0,\mathbf{q}\rightarrow\mathbf{0}),\\ &
&\int_{-\infty}^{+\infty}d\omega\omega\chi_{\rho\rho}(\omega,\mathbf{q})=\frac{nq^2}{m},
\end{eqnarray}
where $\chi_{\rho\rho}=-\frac{1}{\pi}\textrm{Im}Q^{00}_0$ is the density susceptibility and the analytical continuation $i\Omega_l\rightarrow\omega+i0^{+}$
has been applied.
Although these two sum rules can be directly proven by the explicit expressions of the response functions
given in Appendix~\ref{app:a0}, here we give a more instructive proof that has a nice connection with the U(1) symmetries in the Hamiltonian. For the compressibility sum rule, we have
\begin{eqnarray}\label{SRFD1} \frac{\partial n}{\partial \mu}=2\sum_P\frac{\partial
G_0(P)}{\partial \mu}=-2\sum_PG^2_0(P)\frac{\partial G^{-1}_0(P)}{\partial
\mu}=-2\lim_{\mathbf{q}\rightarrow\mathbf{0}}\sum_PG_0(P+Q)G_0(P)|_{\omega=0}=-Q^{00}_0(\omega=0,\mathbf{q}\rightarrow\mathbf{0}),
\end{eqnarray} where the expression (\ref{Qmn0}) and $\gamma^0(P+Q,P)=1$ have been
applied. In fact, this is a special case of the $Q$-limit WI $\Gamma^0=1-(\partial \Sigma/\partial \mu)$, where $\Gamma^0$ and $\gamma^0$ are the full and bare vertex functions, $\Sigma$ is the self energy of fermions, and the limit $\omega=0, q\rightarrow 0$ has been taken. For non-interacting Fermi gases $\Sigma=0$ and $\Gamma^0=\gamma^0=1$ so the $Q$-limit WI is trivially satisfied.
For the $f$-sum rule, we need to implement the real time formalism to describe
the non-equilibrium transport theory. The real time response function corresponding
to Eq.(\ref{Qmn}) is \begin{eqnarray}
Q^{\mu\nu}_0(t-t',\mathbf{q})=-i\theta(t-t')\langle
[J^{\mu}(t,\mathbf{q}),J^{\nu}(t',-\mathbf{q})]\rangle, \end{eqnarray} where the
time-dependent Heisenberg operator is defined by
$\mathcal{O}(t)=e^{iHt}\mathcal{O}e^{-iHt}$. The imaginary part of $Q^{\mu\nu}_0$ is
\begin{eqnarray} \textrm{Im}Q^{\mu\nu}_0(t-t',\mathbf{q})=-\frac{1}{2}\langle
[J^{\mu}(t,\mathbf{q}),J^{\nu}(t',-\mathbf{q})]\rangle. \end{eqnarray} Using
$\omega\textrm{Im}Q^{\mu\nu}_0=\textrm{Im}(\omega Q^{\mu\nu}_0)$, we have
\begin{eqnarray} &
&-\frac{1}{\pi}\int^{+\infty}_{-\infty}d\omega\omega\textrm{Im}Q^{\mu\nu}_0(\omega,\mathbf{q})=\langle
[i\frac{\partial n(t,\mathbf{q})}{\partial
t},n(t,-\mathbf{q})\rangle_{\omega}=\langle
[\mathbf{q}\cdot\mathbf{J}(t,\mathbf{q}),n(t,-\mathbf{q})\rangle_{\omega}\nonumber\\
&=&\sum_{\mathbf{p}\mathbf{p}'}\mathbf{q}\cdot\frac{\mathbf{p}+\frac{\mathbf{q}}{2}}{m}\langle
e^{iHt}\big(\psi^{\dagger}_{\sigma\mathbf{p}}\{\psi_{\sigma\mathbf{p}+\mathbf{q}},\psi^{\dagger}_{\sigma'\mathbf{p}'+\mathbf{q}}\}\psi_{\sigma'\mathbf{p}'}
-\psi^{\dagger}_{\sigma'\mathbf{p}'+\mathbf{q}}\{\psi^{\dagger}_{\sigma\mathbf{p}},\psi_{\sigma'\mathbf{p}'}\}\psi_{\sigma\mathbf{p}+\mathbf{q}}\big)e^{-iHt}\rangle_{\omega}\nonumber\\
&=&\sum_{\mathbf{p}}\mathbf{q}\cdot\frac{(\mathbf{p}+\frac{\mathbf{q}}{2})-(\mathbf{p}-\frac{\mathbf{q}}{2})}{m}\langle
\psi^{\dagger}_{\sigma\mathbf{p}}(t)\psi_{\sigma\mathbf{p}}(t)\rangle=\frac{nq^2}{m},
 \end{eqnarray} where in the first line we used the conservation of
the current $\partial_{\mu}J^{\mu}=0$, in the last line we changed the variables
$\mathbf{p}\rightarrow\mathbf{p}-\mathbf{q}$ for the second term. The subscript
$\omega$ means the Fourier transform of the function with the argument $\omega$.
One thus see that the two sum rules are closely connected to the U(1)$_{\textrm{EM}}$ symmetry.

\section{Gauge-invariant linear response theory from the CFOP approach for Fermionic
Superfluids}\label{C4}
In BCS theory of fermionic superfluids, the order parameter is given by \begin{equation}\label{Delta}
\Delta(\mathbf{x})=g\langle\psi_{\downarrow}(\mathbf{x})\psi_{\uparrow}(\mathbf{x})\rangle.
\end{equation}
Following the mean-field BCS approximation, the Hamiltonian without any external field becomes
\begin{equation}\label{HBCS}
H=\int
d^3\mathbf{\mathbf{x}}\psi^{\dagger}_{\sigma}(\mathbf{x})\Big(\frac{\hat{\mathbf{p}}^2}{2m}
-\mu
\Big)\psi_{\sigma}(\mathbf{x})-\int
d^3\mathbf{\mathbf{x}}\Big(\Delta(\mathbf{x})\psi^{\dagger}_{\uparrow}(\mathbf{x})
\psi^{\dagger
}_{\downarrow}(\mathbf{x})+\textrm{h.c.}\Big).
\end{equation}
One may see that the U(1)$_{\textrm{EM}}$ symmetry is spontaneously
broken if $\Delta(\mathbf{x})\neq 0$ below $T_c$ while the $U(1)_{z}$ symmetry remains intact. Hence the original Hamiltonian (\ref{H00})
has a U(1)$\times$U(1) symmetry while the BCS Hamiltonian (\ref{HBCS}) in the
broken-symmetry phase only has a U(1) symmetry. The phase with a broken
U(1)$_{\textrm{EM}}$ symmetry below $T_c$ brings challenges of how to cast its
associated linear response theory in a gauge invariant form. Below $T_c$, the breaking
of U(1)$_{\textrm{EM}}$ and the unbroken U(1)$_z$ symmetry may be called a ``spin-charge
separation" in BCS theory. We will explore this phenomenon in depth after we present the consistent
linear response theories in both density and spin channels.
Above $T_c$, both symmetries are not broken and the charge
and spin degrees of freedom are not separated there as they do below $T_c$.

Here we present a U(1)$_{\textrm{EM}}$
gauge-invariant linear response theory for BCS superfluids based on the consistent
fluctuation of the order-parameter (CFOP) approach.
The gauge invariance is basically a one-particle problem from the point of
view of quantum field theory while BCS theory is essentially a many-body theory (for
a review, see \cite{Schrieffer_book}). This contrast highlights the importance of Ward identity (or its generalized form)
since one can check the gauge invariance of a theory by verifying the corresponding WI and we will give some
concrete examples.

\subsection{Nambu Based Notation for Linear
Response}
\label{CS2}
As we have seen previously, there are terminologies like the Nambu space, one-dimensional space and the ``Ward identity''(WI) associated
with different physical quantities. Here we explain them in details. The Nambu space is the space in which Nambu spinors are defined. For non-relativistic theories, it is a two-dimensional space and operators are written as two by two matrices.
The basic framework of the CFOP linear response theory for BCS superfluids is easier to develop in the Nambu space.
The terminology one-dimensional space is the abbreviation of the one-dimensional representation space of fermion operators.
The representation space of fermion operators is actually the four-dimensional spinor representation space of the Lorentz
group. In the non-relativistic limit one may approximate it by a one-dimensional space representation if the spin is simply labeled as a subscript. Most linear response theories for normal Fermi gases are formulated in this space since the corrections to the interacting vertex between fermions and external fields are hard to be presented in the Nambu space.
The Ward identity in quantum field theories refers to the relation among the response
(correlation) functions that describe the effects of symmetry
transformations allowed by the theory \cite{CFT}.
Here we use the Ward identity specifically for a diagrammatic identity
between the vertex function and the fermion propagator, which reflects the symmetry from the EM gauge transformation.
 In the Nambu space, we will adopt Nambu's convention to
characterize each of these identities as a ``generalized Ward identity'' (GWI).
There are seminal reviews \cite{Schrieffer_book,Walecka} on how to cast BCS theory in the Nambu space.
However, there are less reviews on the same approach in the one-dimensional space \cite{OurAnnPhys}. A fully gauge invariant EM interacting vertex in the one-dimensional space has not been studied systematically.
Later on, we will present one by using the results found in the Nambu space. Here we first formulate the linear response theory in the
Nambu space for BCS superfluids based on the CFOP approach.

We define
$\sigma_{\pm}=\frac{1}{2}(\sigma_1\pm i\sigma_2)$ in the Nambu space and introduce
the Nambu-Gorkov spinors \begin{equation}\label{Ns}
\Psi_{\mathbf{p}}=\left[\begin{array}{c} \psi_{\uparrow\mathbf{p}} \\
\psi^{\dagger}_{\downarrow-\mathbf{p}}\end{array}\right], \qquad
\Psi^{\dagger}_{\mathbf{p}}=[\psi^{\dagger}_{\uparrow\mathbf{p}},\psi_{\downarrow-\mathbf{p}}].
\end{equation}
Here $\sigma_i$, $i=1,2,3$ are the Pauli matrices.
The Hamiltonian (\ref{EMH}) in
momentum space after the mean-field BCS approximation becomes \begin{eqnarray}\label{H0}
H=\sum_{\mathbf{p}}\Psi^{\dagger}_{\mathbf{p}}\xi_{\mathbf{p}}\sigma_3\Psi_{\mathbf{p}}+\sum_{\mathbf{p}\mathbf{q}}
\Psi^{\dagger}_{\mathbf{p}+\mathbf{q}}\big(-\frac{\mathbf{p}+\frac{\mathbf{q}}{2}}{m}\mathbf{A}_{\mathbf{q}}+\Phi_{\mathbf{q}}\sigma_3-\Delta_{\mathbf{q}}\sigma_+-\Delta^*_{-\mathbf{q}}\sigma_-\big)\Psi_{\mathbf{p}},
\end{eqnarray}
The order parameter (\ref{Delta}) in momentum space is generalized to include finite-momentum fluctuations from its equilibrium value so
$\Delta_{\mathbf{q}}=g\sum_{\mathbf{p}}\langle\Psi^{\dagger}_{\mathbf{p}}\sigma_-\Psi_{\mathbf{p}+\mathbf{q}}\rangle$,
which will be imposed as a self-consistency condition. When the external EM field is
weak, the order parameter is perturbed and deviates from its equilibrium value.
The order parameter in equilibrium is $\Delta$, which is at $\mathbf{q}=\mathbf{0}$ and can be chosen to be real by the U(1)$_{\textrm{EM}}$ symmetry. We denote the
small perturbation of the order parameter as $\Delta'_{\mathbf{q}}$ so
$\Delta_{\mathbf{q}}=\Delta+\Delta'_{\mathbf{q}}$. By introducing
$\Delta_{1\mathbf{q}}=-(\Delta'_{\mathbf{q}}+\Delta^{\prime\ast}_{-\mathbf{q}})/2$
and
$\Delta_{2\mathbf{q}}=-i(\Delta'_{\mathbf{q}}-\Delta^{\prime\ast}_{-\mathbf{q}})/2$,
the Hamiltonian (\ref{H0}) splits into two parts as $H=H_0+H'$ with one containing
the equilibrium quantities and the other containing the deviation from the
equilibrium. \begin{eqnarray}\label{H1}
H_0=\sum_{\mathbf{p}}\Psi^{\dagger}_{\mathbf{p}}\hat{E}_{\mathbf{p}}\Psi_{\mathbf{p}},
\quad
H'=\sum_{\mathbf{p}\mathbf{q}}\Psi^{\dagger}_{\mathbf{p}+\mathbf{q}}\big(\Delta_{1\mathbf{q}}\sigma_1+\Delta_{2\mathbf{q}}\sigma_2+A_{\mu\mathbf{q}}\hat{\gamma}^{\mu}(\mathbf{p}+\mathbf{q},\mathbf{p})\big)\Psi_{\mathbf{p}},
\end{eqnarray}
where $\hat{E}_{\mathbf{p}}=\xi_{\mathbf{p}}\sigma_3-\Delta\sigma_1$
is an energy operator and
$\hat{\gamma}^{\mu}(\mathbf{p}+\mathbf{q},\mathbf{p})=(\sigma_3,\frac{\mathbf{p}+\frac{\mathbf{q}}{2}}{m})$
 is the bare EM vertex in the Nambu space. As discussed in Ref.~\cite{KulikJLTP81}, the fluctuations $\Delta_{1\mathbf{q}}$ and $\Delta_{2\mathbf{q}}$ introduce the amplitude mode and phase mode as the collective modes introduced by the perturbation of the order parameter.

For the equilibrium par, the quasi-particle energy is given by
$E_{\mathbf{p}}=\sqrt{\xi^2_{\mathbf{p}}+\Delta^2}$. The propagator in the Nambu space is
\begin{eqnarray}\label{NGP}
\hat{G}(P)\equiv\hat{G}_{\mathbf{p}}(i\omega_n)=\frac{1}{i\omega_n-\hat{E}_{\mathbf{p}}}=\left(\begin{array}{cc}G(P)
& F(P)\\F(P) & -G(-P)\end{array}\right), \end{eqnarray}
where
\begin{eqnarray}\label{Green}
G(P)=\frac{u^2_{\mathbf{p}}}{i\omega_n-E_{\mathbf{p}}}+\frac{v^2_{\mathbf{p}}}{i\omega_n+E_{\mathbf{p}}},\quad
F(P)=-u_{\mathbf{p}}v_{\mathbf{p}}\Big(\frac{1}{i\omega_n-E_{\mathbf{p}}}-\frac{1}{i\omega_n+E_{\mathbf{p}}}\Big)
\end{eqnarray} are the single-particle Green's function and anomalous Green's
function respectively and
$u^2_{\mathbf{p}},v^2_{\mathbf{p}}=\frac{1}{2}(1\pm\frac{\xi_{\mathbf{p}}}{E_{\mathbf{p}}})$.

The number density and the order parameter in equilibrium can be expressed by the propagator as
\begin{eqnarray}\label{ndNambu} n=\textrm{Tr}\sum_P\big(\sigma_3\hat{G}(P)\big),\quad
\Delta=g\textrm{Tr}\sum_P\big(\sigma_1\hat{G}(P)\big), \end{eqnarray} which give the
number and gap equations \begin{eqnarray}\label{NGE}
n=\sum_{\mathbf{p}}\Big[1-\frac{\xi_{\mathbf{p}}}{E_{\mathbf{p}}}\big(1-2f(E_{\mathbf{p}})\big)\Big],
\quad \frac{1}{g}&=&\sum_{\mathbf{p}}\frac{1-2f(E_{\mathbf{p}})}{2E_{\mathbf{p}}}.
\end{eqnarray} We can also introduce the counterpart of the coefficients $u,v$ in the
Nambu space as
$\hat{u}_{\mathbf{p}},\hat{v}_{\mathbf{p}}=\frac{1}{2}(1\pm\frac{\hat{E}_{\mathbf{p}}}{E_{\mathbf{p}}})$.
Using the properties $\hat{u}^2_{\mathbf{p}}=\hat{u}_{\mathbf{p}}$,
$\hat{v}^2_{\mathbf{p}}=\hat{v}_{\mathbf{p}}$ and
$\hat{u}_{\mathbf{p}}\hat{v}_{\mathbf{p}}=0$, one can show that in the Nambu space the
propagator has the form \begin{eqnarray}\label{G2}
\hat{G}(P)=\frac{\hat{u}^2_{\mathbf{p}}}{i\omega_n-E_{\mathbf{p}}}+\frac{\hat{v}^2_{\mathbf{p}}}{i\omega_n+E_{\mathbf{p}}}.
\end{eqnarray} This has a similar expression as the single particle Green's function shown in Eq.~\eqref{Green}.
In fact, any function of $i\omega_n-\hat{E}_{\mathbf{p}}$,
for example $F(i\omega_n-\hat{E}_{\mathbf{p}})$, can be expressed as $
F(i\omega_n-\hat{E}_{\mathbf{p}})=\hat{u}_{\mathbf{p}}(i\omega_n-E_{\mathbf{p}})+\hat{v}_{\mathbf{p}}(i\omega_n+E_{\mathbf{p}})$ \cite{OurPRD12}.
It is an interesting property of BCS theory in the Nambu space and will be useful for deriving the response functions.

The interacting Hamiltonian $H'$ in
Eqs.~(\ref{H1}) can be cast in the form
\begin{eqnarray}\label{H2}
H'=\sum_{\mathbf{p}\mathbf{q}}\Psi^{\dagger}_{\mathbf{p}+\mathbf{q}}\hat{\Phi}^T_{\mathbf{q}}\cdot\hat{\Sigma}(\mathbf{p}+\mathbf{q},\mathbf{p})\Psi_{\mathbf{p}},
\end{eqnarray} where \begin{eqnarray}\label{GGP}
\hat{\mathbf{\Phi}}_{\mathbf{q}}=\big(\Delta_{1\mathbf{q}},\Delta_{2\mathbf{q}},A_{\mu\mathbf{q}}\big)^T,\quad
\hat{\mathbf{\Sigma}}(\mathbf{p}+\mathbf{q},\mathbf{p})=\big(\sigma_1,\sigma_2,\hat{\gamma}^{\mu}(\mathbf{p}+\mathbf{q},\mathbf{p})\big)^T,
\end{eqnarray}
are defined as the generalized driving potential and generalized
interacting vertex, respectively. By using the imaginary time formalism,
we assume that under the perturbation from $H'$, the generalized perturbed current
$\delta \vec{J}$ is given by \begin{eqnarray}\label{RF1}
\delta \vec{J}(\tau,\mathbf{q})=\sum_{\mathbf{p}}\langle\Psi^{\dagger}_{\mathbf{p}}(\tau)\hat{\mathbf{\Sigma}}(\mathbf{p}+\mathbf{q},\mathbf{p})\Psi_{\mathbf{p}+\mathbf{q}}(\tau)\rangle+\frac{n}{m}\delta^{i3}h^{\mu\nu}A_{\nu}(\tau,\mathbf{q}).
\end{eqnarray} Here $\delta J^{\mu}_3$ denotes the perturbed EM
current and $\delta J_{1,2}$ denote the perturbations of the gap function. The linear
response theory is written in a matrix form
\begin{eqnarray}\label{eqn:Q}
\delta \vec{J}(\omega,\mathbf{q})&=&\tensor{Q}(\omega,\mathbf{q})\cdot\hat{\mathbf{\Phi}}(\omega,\mathbf{q})
\nonumber \\ &=&\left( \begin{array}{ccc} Q_{11}(\omega,\mathbf{q}) &
Q_{12}(\omega,\mathbf{q}) & Q^{\nu}_{13}(\omega,\mathbf{q}) \\
Q_{21}(\omega,\mathbf{q}) & Q_{22}(\omega,\mathbf{q}) &
Q^{\nu}_{23}(\omega,\mathbf{q}) \\ Q^{\mu}_{31}(\omega,\mathbf{q}) &
Q^{\mu}_{32}(\omega,\mathbf{q}) &
Q^{\mu\nu}_{33}(\omega,\mathbf{q})+\frac{n}{m}h^{\mu\nu} \end{array}\right)\left(
\begin{array}{ccc} \Delta_{1}(\omega,\mathbf{q}) \\ \Delta_{2}(\omega,\mathbf{q}) \\A_{\nu}(\omega,\mathbf{q})
\end{array}\right).
\end{eqnarray}
The response functions $Q_{ij}$ are
\begin{eqnarray} Q_{ij}(\tau-\tau',\mathbf{q})=-\sum_{\mathbf{p}\mathbf{p}'}\langle
T_{\tau}[\Psi^{\dagger}_{\mathbf{p}}(\tau)\hat{\Sigma}_i(\mathbf{p}+\mathbf{q},\mathbf{p})\Psi_{\mathbf{p}+\mathbf{q}}(\tau)\Psi^{\dagger}_{\mathbf{p}'+\mathbf{q}}(\tau')\hat{\Sigma}_j(\mathbf{p}',\mathbf{p}'+\mathbf{q})\Psi_{\mathbf{p}'}(\tau')]\rangle,
\end{eqnarray} After applying Fourier transform and Wick's theorem, we obtain
\begin{eqnarray}\label{eqn:RF} Q_{ij}(i\Omega_{l},
\mathbf{q})=\textrm{Tr}T\sum_{i\omega_n}\sum_{\mathbf{p}}
\big(\hat{\Sigma}_i(P+Q,P)\hat{G}(P+Q)\hat{\Sigma}_j(P,P+Q)\hat{G}(P)\big),
\end{eqnarray} For convenience, we have defined
$\hat{\gamma}^{\mu}(P+Q,P)\equiv\hat{\gamma}^{\mu}(\mathbf{p}+\mathbf{q},\mathbf{p})$, hence
$\hat{\mathbf{\Sigma}}(P+Q,P)=\hat{\mathbf{\Sigma}}(\mathbf{p}+\mathbf{q},\mathbf{p})$.
By using the expression (\ref{G2}) the response functions are evaluated and shown in Appendix~\ref{app:a0}.

The perturbation of the order parameter and the EM perturbation are treated on
equal footing and this will
naturally lead to the gauge invariance of the CFOP linear response theory. The gap
equation gives the self-consistent condition $\delta J_{1,2}=-\frac{2}{g}\Delta_{1,2}$. 
Applying this relation to Eq.(\ref{eqn:Q}), we get \begin{eqnarray}\label{D1D2}
\Delta_1=-\frac{Q^{\nu}_{13}\tilde{Q}_{22}-Q^{\nu}_{23}Q_{12}}{\tilde{Q}_{11}\tilde{Q}_{22}-Q_{12}Q_{21}}A_{\nu},\quad\Delta_2=-\frac{Q^{\nu}_{23}\tilde{Q}_{11}-Q^{\nu}_{13}Q_{21}}{\tilde{Q}_{11}\tilde{Q}_{22}-Q_{12}Q_{21}}A_{\nu}.
\end{eqnarray} where $\tilde{Q}_{11}\equiv \frac{2}{g}+Q_{11}$ and
$\tilde{Q}_{22}\equiv \frac{2}{g}+Q_{22}$. After substituting the results into
\begin{eqnarray}\label{CEtemp}
\delta J^{\mu}=Q^{\mu}_{31}\Delta_1+Q^{\mu}_{32}\Delta_2+(Q^{\mu\nu}_{33}+\frac{n}{m}h^{\mu\nu})A_{\nu},
\end{eqnarray} we get $\delta J^{\mu}=K^{\mu\nu}A_{\nu}$, where the corrected EM response
kernel $K^{\mu\nu}$ including the effects of fluctuations of the order parameter is
given by \begin{eqnarray}\label{dK} K^{\mu\nu}=\tilde{Q}^{\mu\nu}_{33}+\delta
K^{\mu\nu},\quad \delta
K^{\mu\nu}=-\frac{\tilde{Q}_{11}Q^{\mu}_{32}Q^{\nu}_{23}+\tilde{Q}_{22}Q^{\mu}_{31}Q^{\nu}_{13}-Q_{12}Q^{\mu}_{31}Q^{\nu}_{23}-Q_{21}Q^{\mu}_{32}Q^{\nu}_{13}}{\tilde{Q}_{11}\tilde{Q}_{22}-Q_{12}Q_{21}}.
\end{eqnarray} Here $\tilde{Q}^{\mu\nu}_{33}=Q^{\mu\nu}_{33}+\frac{n}{m}h^{\mu\nu}$.
The gauge invariance condition for the perturbed current, $q_{\mu} \delta J^{\mu}=0$, is
satisfied if $q_{\mu}K^{\mu\nu}(Q)=0$. This is further guaranteed by the following WIs for the
response functions \begin{eqnarray}\label{WI} q_{\mu}Q^{\mu}_{31}=-2i\Delta
Q_{21},\quad q_{\mu}Q^{\mu}_{32}=-2i\Delta\tilde{Q}_{22},\quad
q_{\mu}\tilde{Q}^{\mu\nu}_{33}=-2i\Delta Q^{\nu}_{23}. \end{eqnarray}
These relations
are more complicated than $q_{\mu}\tilde{Q}^{\mu\nu}_0(Q)=0$ for noninteracting Fermi gases since the perturbation of the order parameter
comes into the linear response theory. The gauge invariance condition of the
perturbed EM current is immediately derived from Eq.~(\ref{WI}) \begin{eqnarray}
q_{\mu}K^{\mu\nu}&=&-2i\Delta
Q^{\nu}_{23}+2i\Delta\frac{\tilde{Q}_{11}\tilde{Q}_{22}Q^{\nu}_{23}+\tilde{Q}_{22}Q_{21}Q^{\nu}_{13}-Q_{12}Q_{21}Q^{\nu}_{23}-Q_{21}\tilde{Q}_{22}Q^{\nu}_{13}}{\tilde{Q}_{11}\tilde{Q}_{22}-Q_{12}Q_{21}}\nonumber\\
&=&-2i\Delta Q^{\nu}_{23}+2i\Delta Q^{\nu}_{23}=0. \end{eqnarray}

Now we sketch the
proof of the WIs (\ref{WI}). The bare inverse propagator in the Nambu space is
$\hat{G}^{-1}_0(P)=i\omega_n-\xi_{\mathbf{p}}\sigma_3$ and the self energy
is $\hat{\Sigma}=-\Delta\sigma_1$ \cite{Schrieffer_book}. Hence from Eq.(\ref{NGP}) the full inverse
propagator can be expressed as $\hat{G}^{-1}(P)=\hat{G}^{-1}_0(P)-\hat{\Sigma}$, and
one can verify that \begin{eqnarray}\label{WI0}
\sigma_3\hat{G}^{-1}(P+Q)-\hat{G}^{-1}(P)\sigma_3=i\Omega_l\sigma_3-(\xi_{\mathbf{p}+\mathbf{q}}-\xi_{\mathbf{p}})+2i\Delta\sigma_2=q_{\mu}\hat{\gamma}^{\mu}(P+Q,P)+2i\Delta\sigma_2.
\end{eqnarray} In fact, as we will point out later, this identity is actually the GWI
that connects the full EM interacting vertex and the full propagator in the Nambu space.
Now we show how the WIs are derived from Eq.(\ref{WI0}).
For the first identity of the WIs, we have \begin{eqnarray}\label{WI1}
q_{\mu}Q^{\mu}_{31}+2i\Delta
Q_{21}
&=&\textrm{Tr}\sum_P\Big[\big(\sigma_3\hat{G}^{-1}(P+Q)-\hat{G}^{-1}(P)\sigma_3\big)\hat{G}(P+Q)\sigma_1\hat{G}(P)\Big]\nonumber\\
&=&\textrm{Tr}\sum_P\big[i\sigma_2\hat{G}(P)\big]+\textrm{Tr}\sum_P\big[\hat{G}(P+Q)i\sigma_2\big]=2\textrm{Tr}\sum_P\big[i\sigma_2\hat{G}(P)\big]=0, \end{eqnarray}
where the cyclic property of the trace has been applied. For the second identity of
WIs, we have \begin{eqnarray}\label{WI2} q_{\mu}Q^{\mu}_{32}+2i\Delta
Q_{22}
&=&\textrm{Tr}\sum_P\Big[\big(\sigma_3\hat{G}^{-1}(P+Q)-\hat{G}^{-1}(P)\sigma_3\big)\hat{G}(P+Q)\sigma_2\hat{G}(P)\Big]\nonumber\\
&=&-2\textrm{Tr}\sum_P\big[i\sigma_1\hat{G}(P)\big]=-4i\textrm{Tr}\sum_PF(P)=-4i\frac{\Delta}{g}. \end{eqnarray}
Therefore $q_{\mu}Q^{\mu}_{32}=-2i\Delta(Q_{22}+\frac{2}{g})=-2i\Delta\tilde{Q}_{22}$
and we get the second WI for the response functions. For the last WI, we have
\begin{eqnarray}\label{WI3} q_{\mu}\tilde{Q}^{\mu\nu}_{33}+2i\Delta
Q^{\nu}_{23}
&=&\textrm{Tr}\sum_P\Big[\sigma_3\hat{\gamma}^{\nu}(P+Q,P)\hat{G}(P)\Big]-\textrm{Tr}\sum_P\Big[\hat{G}(P+Q)\hat{\gamma}^{\nu}(P,P+Q)\sigma_3\Big]-\frac{n}{m}q^{\nu}(1-\eta^{\nu0})\nonumber\\
&=&\sum_{P}\textrm{Tr}\big(\hat{G}(P)\sigma_3[\hat{\gamma}^{\nu}(P+Q,P)-\hat{\gamma}^{\nu}(P-Q,P)]\big)-\frac{n}{m}q^{\nu}(1-\eta^{\nu0})\nonumber\\
&=&\frac{q^{\nu}}{m}(1-\eta^{\nu0})\sum_{P}\textrm{Tr}\big(\hat{G}(P)\sigma_3\big)-\frac{n}{m}q^{\nu}(1-\eta^{\nu0})=0,
 \end{eqnarray} where the fact that $\sigma_3$ commutes with
$\hat{\gamma}^{\mu}$ has been applied and the number
equation (\ref{ndNambu}) has been used.

The CFOP linear response theory also satisfies the $f$-sum rule
\begin{eqnarray}\label{fs}
\int_{-\infty}^{+\infty}d\omega\omega\chi_{\rho\rho}(\omega,\mathbf{q})=n\frac{q^2}{m}.
\end{eqnarray}
Here $\chi_{\rho\rho}=-\frac{1}{\pi}\textrm{Im}K^{00}$ with $K^{00}$ given
by the $00$-component of Eq.~(\ref{dK}):
\begin{eqnarray}\label{K00}
K^{00}=\tilde{Q}^{00}_{33}-\frac{\tilde{Q}_{11}Q^{0}_{32}Q^{0}_{23}+\tilde{Q}_{22}Q^{0}_{31}Q^{0}_{13}-Q_{12}Q^{0}_{31}Q^{0}_{23}-Q_{21}Q^{0}_{32}Q^{0}_{13}}{\tilde{Q}_{11}\tilde{Q}_{22}-Q_{12}Q_{21}}.
\end{eqnarray} The following lemma will be useful for the proof of the $f$-sum rule
\begin{eqnarray}\label{fsl1}
-\int_{-\infty}^{+\infty}d\omega\frac{1}{\pi}\textrm{Im}\big[\mathbf{q}\cdot\mathbf{Q}^0_{33}(\omega,\mathbf{q})\big]=n\frac{q^2}{m}.
\end{eqnarray}
The proof of this lemma can be found in Appendix~\ref{app:a1}.
From $\omega\textrm{Im}K^{00}=\textrm{Im}(\omega
K^{00})$ and the third equation of the WIs (\ref{WI}), we have
\begin{eqnarray} \omega K^{00}&=&\mathbf{q}\cdot\mathbf{Q}^0_{33}-2i\Delta
Q^0_{23}-\frac{1}{\tilde{Q}_{11}\tilde{Q}_{22}-Q_{12}Q_{21}}\nonumber\\ &
&\times\big(\tilde{Q}_{11}\mathbf{q}\cdot\mathbf{Q}_{32}Q^0_{23}-2i\Delta\tilde{Q}_{11}\tilde{Q}_{22}Q^0_{23}+\tilde{Q}_{22}\mathbf{q}\cdot\mathbf{Q}_{31}Q^0_{13}-2i\Delta\tilde{Q}_{22}Q_{21}Q^0_{13}\nonumber\\
& &-Q_{12}\mathbf{q}\cdot\mathbf{Q}_{31}Q^0_{23}+2i\Delta Q_{12}Q_{21}Q^0_{23}
-Q_{21}\mathbf{q}\cdot\mathbf{Q}_{32}Q^0_{13}+2i\Delta
Q_{21}\tilde{Q}_{22}Q^0_{13}\big). \end{eqnarray} Note $Q_{12}=-Q_{21}$,
$Q^0_{23}=-Q^0_{32}$, $\mathbf{Q}_{23}=-\mathbf{Q}_{32}$, $Q^0_{13}=Q^0_{31}$ and
$\mathbf{Q}_{13}=\mathbf{Q}_{31}$, we have \begin{eqnarray} \omega
K^{00}&=&\mathbf{q}\cdot\mathbf{Q}^0_{33}-\frac{\tilde{Q}_{11}\mathbf{q}\cdot\mathbf{Q}_{32}Q^0_{23}+\tilde{Q}_{22}\mathbf{q}\cdot\mathbf{Q}_{31}Q^0_{13}-Q_{12}\mathbf{q}\cdot\mathbf{Q}_{31}Q^0_{23}-Q_{21}\mathbf{q}\cdot\mathbf{Q}_{32}Q^0_{13}}{\tilde{Q}_{11}\tilde{Q}_{22}-Q_{12}Q_{21}}.
\end{eqnarray} Using the lemma, it can be shown that proving the $f$-sum rule is
equivalent to proving
\begin{eqnarray}
\int_{-\infty}^{+\infty}d\omega\textrm{Im}\frac{\tilde{Q}_{11}\mathbf{q}\cdot\mathbf{Q}_{32}Q^0_{23}+\tilde{Q}_{22}\mathbf{q}\cdot\mathbf{Q}_{31}Q^0_{13}-Q_{12}\mathbf{q}\cdot\mathbf{Q}_{31}Q^0_{23}-Q_{21}\mathbf{q}\cdot\mathbf{Q}_{32}Q^0_{13}}{\tilde{Q}_{11}\tilde{Q}_{22}-Q_{12}Q_{21}}=0.
\end{eqnarray}
Note that $\tilde{Q}_{11}$, $\tilde{Q}_{22}$, $Q^0_{13}$ and
$\mathbf{Q}_{23}$ are even functions of $\omega$, while $Q_{12}$, $\mathbf{Q}_{13}$
and $Q^0_{23}$ are odd functions of $\omega$. Hence the integrand is an odd function
of $\omega$, then the integral indeed vanishes. Therefore, the $f$-sum rule (\ref{fs}) is satisfied by the density
linear response theory. From the proof we can see that we need the explicit expressions of the
response functions. In fact we can not prove the $f$-sum rule simply by using the WIs
(\ref{WI}) or the conservation law of current $\partial_{\mu}J^{\mu}=0$ alone, which is different from what we
have done in the case of non-interacting Fermi gases.

\subsection{WI and $Q$-limit WI of the CFOP Linear Response Theory of BCS
Superfluids}
\label{CS3}

\subsubsection{GWI and $Q$-limit GWI in the Nambu space}\label{CS4}
To complete our
discussions on the CFOP theory in the Nambu space, we now investigate the GWI which
connects the full EM vertex and the fermion propagator.
Due to different representations for the fermions, we will present the GWI in the Nambu space as well as
the WI in the one-dimensional space
One can verify that the bare EM vertex
and the propagator given in Sec.~\ref{CS2} satisfies the bare GWI in the Nambu space
\begin{eqnarray}\label{bGWI}
q_{\mu}\hat{\gamma}^{\mu}(P+Q,P)=\sigma_3\hat{G}^{-1}_0(P+Q)-\hat{G}^{-1}_0(P)\sigma_3.
\end{eqnarray}
Hence, as pointed by Nambu \cite{Nambu60}, the GWI for full EM vertex has the similar
structure \begin{equation}\label{GWI5}
q_{\mu}\hat{\Gamma}^{\mu}(P+Q,P)=\sigma_3\hat{G}^{-1}(P+Q)-\hat{G}^{-1}(P)\sigma_3,
\end{equation}
Moreover, the corrected EM response kernel $K^{\mu\nu}$ given by
Eq.~(\ref{dK}) should be evaluated by the full EM vertex and propagator in the Nambu space as
\begin{eqnarray}\label{Kmn1}
K^{\mu\nu}(Q)=\textrm{Tr}\sum_P\big(\hat{\Gamma}^{\mu}(P+Q,P)\hat{G}(P+Q)\hat{\gamma}^{\nu}(P,P+Q)\hat{G}(P)\big)+\frac{n}{m}h^{\mu\nu},
\end{eqnarray}
a similar form in the one dimensional space will be presented later in
Eq.(\ref{KmnO}). In fact, this full EM vertex $\hat{\Gamma}^{\mu}(P+Q,P)$ can be
inferred from the expression of $K^{\mu\nu}$. We define
\begin{eqnarray}\label{tmp4}
\Pi^{\mu}_1=\frac{\left|\begin{array}{cc}Q^{\mu}_{31} & Q_{21}\\ Q^{\mu}_{32} &
\tilde{Q}_{22}\end{array}\right|}{\left|\begin{array}{cc}\tilde{Q}_{11} & Q_{12}\\
Q_{21} & \tilde{Q}_{22}\end{array}\right|}, \mbox{
}\Pi^{\mu}_2=\frac{\left|\begin{array}{cc}Q^{\mu}_{32} & Q_{12} \\ Q^{\mu}_{31} &
\tilde{Q}_{11} \end{array}\right|}{\left|\begin{array}{cc}\tilde{Q}_{11} & Q_{12}\\
Q_{21} & \tilde{Q}_{22}\end{array}\right|}. \end{eqnarray}
Hence, from Eq.(\ref{dK}) $K^{\mu\nu}$ can be expressed as
\begin{eqnarray}\label{Kmn2}
K^{\mu\nu}(Q)&=&Q^{\mu\nu}_{33}(Q)-\Pi^{\mu}_1(Q)Q^{\nu}_{13}(Q)-\Pi^{\mu}_2(Q)Q^{\nu}_{23}(Q)+\frac{n}{m}h^{\mu\nu}\nonumber\\
&=&\textrm{Tr}\sum_P\big(\hat{\gamma}^{\mu}(P+Q,P)\hat{G}(P+Q)\hat{\gamma}^{\nu}(P,P+Q)\hat{G}(P)\big)-\Pi^{\mu}_1(Q)\textrm{Tr}\sum_P\big(\sigma_1\hat{G}(P+Q)\hat{\gamma}^{\nu}(P,P+Q)\hat{G}(P)\big)\nonumber\\
&
&-\Pi^{\mu}_2(Q)\textrm{Tr}\sum_P\big(\sigma_2\hat{G}(P+Q)\hat{\gamma}^{\nu}(P,P+Q)\hat{G}(P)\big)+\frac{n}{m}h^{\mu\nu}\nonumber\\
&=&\textrm{Tr}\sum_P\big([\hat{\gamma}^{\mu}(P+Q,P)-\sigma_1\Pi_1(Q)-\sigma_2\Pi_2(Q)]\hat{G}(P+Q)\hat{\gamma}^{\nu}(P,P+Q)\hat{G}(P)\big)+\frac{n}{m}h^{\mu\nu}.
\end{eqnarray}
where in the second line we have substituted the expression
(\ref{eqn:RF}) for $Q^{\mu\nu}_{33}$, $Q^{\nu}_{13}$ and $Q^{\nu}_{23}$. Comparing
Eq.(\ref{Kmn1}) with the last line of Eq.~(\ref{Kmn2}), one can find that the full EM
vertex is given by
\begin{eqnarray}\label{FEM}
\hat{\Gamma}^{\mu}(P+Q,P)=\hat{\gamma}^{\mu}(P+Q,P)-\sigma_1\Pi^{\mu}_1(Q)-\sigma_2\Pi^{\mu}_2(Q).
\end{eqnarray}
By applying the WIs (\ref{WI}), one can see that $\Pi^{\mu}_{1,2}$
satisfies $q_{\mu}\Pi^{\mu}_1(Q)=0$ and $q_{\mu}\Pi^{\mu}_2(Q)=-2i\Delta$. Hence, the
full EM vertex given by Eq.~(\ref{FEM}) further satisfies
\begin{eqnarray}\label{FGWI}
q_{\mu}\hat{\Gamma}^{\mu}(P+Q,P)=q_{\mu}\hat{\gamma}^{\mu}(P+Q,P)+2i\Delta\sigma_2=\sigma_3\hat{G}^{-1}_0(P+Q)-\hat{G}^{-1}_0(P)\sigma_3,
\end{eqnarray}
where Eq.~(\ref{WI0}) as been applied. Therefore the full EM vertex
indeed obeys the important GWI in the Nambu space.

Having the expression of full EM vertex, one can show that it also respects the $Q$-limit GWI
\begin{eqnarray}\label{CSRN0}
\lim_{\mathbf{q}\rightarrow\mathbf{0}}\hat{\Gamma}^0(P+Q,P)|_{\omega=0}=\frac{\partial \hat{G}^{-1}(P)}{\partial \mu}=\sigma_3-\frac{\partial \hat{\Sigma}(P)}{\partial \mu}.
\end{eqnarray}
which is a sufficient and necessary condition for the compressibility sum rule \cite{Maebashi09}.
\begin{eqnarray}
\frac{\partial n}{\partial \mu}=-K^{00}(\omega=0,\mathbf{q}\rightarrow\mathbf{0}).
\end{eqnarray}
This is proven as the following:
\begin{eqnarray}\label{CSRN} \frac{\partial n}{\partial
\mu}&=&\textrm{Tr}\sum_P\Big(\frac{\partial \hat{G}(P)}{\partial
\mu}\sigma_3\Big)=-\textrm{Tr}\sum_P\Big(\hat{G}(P)\big(\sigma_3-\frac{\partial
\hat{\Sigma}(P)}{\partial \mu}\big)\hat{G}(P)\sigma_3\Big)\nonumber\\
&=&-\textrm{Tr}\sum_P\Big(\hat{\Gamma}^0(P,P)\hat{G}(P)\hat{\gamma}^0(P,P)\hat{G}(P)\Big)=-K^{00}(\omega=0,\mathbf{q}\rightarrow\mathbf{0}).
\end{eqnarray}
The $Q$-limit GWI (\ref{CSRN0}) has a profound physical implication for interacting Fermi gases. The LHS of it is associated
with the equation of state derivable from one-particle correlation functions, while the RHS involves the response function evaluated by two-particle correlation functions. Hence
it builds a bridge connecting the one-particle and two-particle formalisms.

There is a subtlety that needs to be addressed here.
It is on whether the $Q$-limit GWI (\ref{CSRN0}) can be derived from the GWI (\ref{GWI5}), or equivalently, whether the $Q$-limit GWI serves as an independent constraint. Here we argue that the GWI imposes no
constraint on the form of the $Q$-limit GWI. The reason is because in the limit $\omega=0$ and $\mathbf{q}\rightarrow\mathbf{0}$, the GWI (\ref{GWI5})
becomes $\mathbf{q}\cdot\hat{\mathbf{\Gamma}}(P,P)=\lim_{\mathbf{q}\rightarrow\mathbf{0}}[\sigma_3\hat{G}^{-1}(P+Q)-\hat{G}^{-1}(P)\sigma_3]|_{\omega=0}$,
 where $\hat{\mathbf{\Gamma}}$ is the spatial component of the vertex function. However,
the $Q$-limit GWI is an identity regarding $\hat{\Gamma}^0$ so the GWI does not reveal any information about the $Q$-limit GWI in the
limit $\omega=0$ and $\mathbf{q}\rightarrow\mathbf{0}$. Thus one should treat the $Q$-limit GWI as an independent constraint of a linear response
theory. The problem of how to obtain a consistent expression for the compressibility as discussed in Ref.~\cite{Mahanbook} thus can be
rephrased as how a linear response theory can satisfy the compressibility sum rule, or more directly whether the $Q$-limit GWI could be satisfied.

Now we show that
the $0$-th component of the $\hat{\Gamma}^{\mu}$ given by Eq.(\ref{FEM}) satisfies the
$Q$-limit GWI. Since $Q_{12}(Q)=-Q_{21}(Q)=Q^0_{23}(Q)=-Q^0_{32}(Q)=0$ when
$\omega=0$, then $\Pi^0_2(Q)=0$ in this limit. Hence we have
\begin{eqnarray}\label{QWIP1}
\lim_{\mathbf{q}\rightarrow\mathbf{0}}\hat{\Gamma}^0(P+Q,P)\Big|_{\omega=0}=\sigma_3-\sigma_1\lim_{\mathbf{q}\rightarrow\mathbf{0}}\Pi_1^0(Q)\Big|_{\omega=0}=\sigma_3-\sigma_1\frac{Q^0_{13}(0,\mathbf{q}\rightarrow\mathbf{0})}{\tilde{Q}_{11}(0,\mathbf{q}\rightarrow\mathbf{0})},
\end{eqnarray} where \begin{eqnarray}\label{Qres}
Q^0_{13}(0,\mathbf{q}\rightarrow\mathbf{0})&=&-\Delta\sum_{\mathbf{p}}\frac{\xi_{\mathbf{p}}}{E^2_{\mathbf{p}}}\Big[\frac{1-2f(E_{\mathbf{p}})}{E_{\mathbf{p}}}+2\frac{\partial
f(E_{\mathbf{p}})}{\partial E_{\mathbf{p}}}\Big],\nonumber\\
\tilde{Q}_{11}(0,\mathbf{q}\rightarrow\mathbf{0})&=&\Delta^2\sum_{\mathbf{p}}\frac{1}{E^2_{\mathbf{p}}}\Big[\frac{1-2f(E_{\mathbf{p}})}{E_{\mathbf{p}}}+2\frac{\partial
f(E_{\mathbf{p}})}{\partial E_{\mathbf{p}}}\Big]. \end{eqnarray} Note the self-energy
operator is given by $\hat{\Sigma}=-\sigma_1\Delta$, hence we have \begin{eqnarray}
\frac{\partial \hat{\Sigma}}{\partial \mu}=-\sigma_1\frac{\partial \Delta}{\partial
\mu}. \end{eqnarray} We need to evaluate $\frac{\partial \Delta}{\partial \mu}$ from
the gap equation (\ref{NGE}). Differentiating both sides with respect
to $\mu$ one gets
\begin{eqnarray}\label{pDpm}
\frac{\partial\Delta}{\partial\mu}=\frac{\sum_{\mathbf{p}}\frac{\xi_{\mathbf{p}}}{E^2_{\mathbf{p}}}\Big(\frac{1-2f(E_{\mathbf{p}})}{E_{\mathbf{p}}}+2\frac{\partial
f(E_{\mathbf{p}})}{\partial
E_{\mathbf{p}}}\Big)}{\sum_{\mathbf{p}}\frac{\Delta}{E^2_{\mathbf{p}}}\Big(\frac{1-2f(E_{\mathbf{p}})}{E_{\mathbf{p}}}+2\frac{\partial
f(E_{\mathbf{p}})}{\partial
E_{\mathbf{p}}}\Big)}=-\frac{Q^0_{13}(0,\mathbf{q}\rightarrow\mathbf{0})}{\tilde{Q}_{11}(0,\mathbf{q}\rightarrow\mathbf{0})}=-\lim_{\mathbf{q}\rightarrow\mathbf{0}}\Pi_1^0(Q)\Big|_{\omega=0}.
\end{eqnarray} Plug this into the right-hand side of Eq.~(\ref{QWIP1}), we prove the $Q$-limit GWI
(\ref{CSRN}) for the full EM vertex.

Here we summarize our key results in the Nambu space:
\begin{itemize} \item Response
kernel: \begin{eqnarray}\label{Kmn3}
    K^{\mu\nu}(Q)=\textrm{Tr}\sum_P\big(\hat{\Gamma}^{\mu}(P+Q,P)\hat{G}(P+Q)\hat{\gamma}^{\nu}(P,P+Q)\hat{G}(P)\big)+\frac{n}{m}h^{\mu\nu},\end{eqnarray}
\item Full EM interacting vertex: \begin{eqnarray}\hat{\Gamma}^{\mu}(P+Q,P)=\hat{\gamma}^{\mu}(P+Q,P)-\sigma_1\Pi^{\mu}_1(Q)-\sigma_2\Pi^{\mu}_2(Q),\end{eqnarray}
\item Generalized Ward identity: \begin{eqnarray}\label{GWI}\sigma_3\hat{G}^{-1}(P+Q)-\hat{G}^{-1}(P)\sigma_3=q_{\mu}\hat{\Gamma}^{\mu}(P+Q,P),\end{eqnarray}
\item $Q$-limit generalized Ward identity:
\begin{eqnarray}\label{QGWI}
\hat{\Gamma}^0(P,P)=\sigma_3-\frac{\partial \hat{\Sigma}(P)}{\partial \mu},
\end{eqnarray}
\item $f$-sum rule:
\begin{eqnarray}\label{fsN}
\int_{-\infty}^{+\infty}d\omega\omega\chi_{\rho\rho}(\omega,\mathbf{q})=n\frac{q^2}{m}.
\end{eqnarray}
\end{itemize}
We emphasize that a consistent linear response theory for BCS superfluids should satisfy the GWI, $Q$-limit GWI, and $f$-sum rule.

\subsubsection{WI and $Q$-limit WI in the one-dimensional space}\label{CS5}
We have shown
the details of the CFOP theory for BCS superfluids in the Nambu space.
Instead of formulating BCS theory in the matrix form in the Nambu space, one may use the Green's functions \eqref{Green} to formulate BCS theory.
The Leggett-BCS theory of BCS-BEC crossover \cite{Leggett,OurReview} follows this path and we call this ``BCS theory in the one-dimensional space".
This representation is convenient when generalized to the theories for BCS-BEC crossover.
One may
build the CFOP theory step by step in the one-dimensional space similar to what we have done so far.
Here we simply extract the key results from those in the Nambu space since the underlying physics is the same.
We only need to
find the full EM vertex $\Gamma^{\mu}$ in the one-dimensional space such that the EM
response kernel can be evaluated by
\begin{eqnarray}\label{KmnO}
K^{\mu\nu}(Q)=2\sum_{P}\Gamma^{\mu}(P+Q,P)G(P+Q)
\gamma^{\nu}(P,P+Q)G(P)+\frac{n}{m}h^{\mu\nu}, \end{eqnarray}
where $\gamma^{\mu}$ is
the bare EM vertex given in Section \ref{C2}. Moreover, the full EM vertex
$\Gamma^{\mu}$ must obey the WI
\begin{eqnarray}\label{IWI}
q_{\mu}\Gamma^{\mu}(P+Q,P)=G^{-1}(P+Q)-G^{-1}(P)
\end{eqnarray}
and the $Q$-limit WI
\begin{eqnarray}\label{QWI}
\lim_{\mathbf{q}\rightarrow\mathbf{0}}\Gamma^0(P+Q,P)|_{\omega=0}=\frac{\partial G^{-1}(P)}{\partial \mu}=1-\frac{\partial \Sigma(P)}{\partial \mu}.
\end{eqnarray}
The expressions of
the gauge-invariant response kernels should not depend on the space in which we evaluate the response functions. Hence $K^{\mu\nu}(Q)$ given by Eq.~(\ref{KmnO}) must be the same as that in
Eq.~(\ref{Kmn1}). We can derive $\Gamma^{\mu}$ by using this relation.
We emphasize that the two
sides of the WIs (\ref{IWI}) have very different meanings. The left-hand side, which involves the interaction between
fermions and the external field, is a single-particle process. The right-hand side, however, contains many
particle effects since the Green's function contains the self energy that represents the interactions among particles.

The bare EM vertex in the Nambu space can be expressed as
$\hat{\gamma}^{\mu}(P+Q,P)=\textrm{diag}(\gamma^{\mu}(P+Q,P),-\gamma^{\mu}(-P,-P-Q))$.
We define \begin{equation}\label{OWI1}
\Pi^{\mu}(Q)=-\Pi^{\mu}_1(Q)+i\Pi^{\mu}_2(Q),\quad
\bar{\Pi}^{\mu}(Q)=-\Pi^{\mu}_1(Q)-i\Pi^{\mu}_2(Q), \end{equation} which satisfy
\begin{equation}\label{OWI2} q_{\mu}\Pi^{\mu}(Q)=2\Delta,\quad
q_{\mu}\bar{\Pi}^{\mu}(Q)=-2\Delta. \end{equation} In fact these two equations are
the off-diagonal terms of the GWI (\ref{GWI5}) in the Nambu space. Hence from Eq.(\ref{FEM})
the matrix form of the full EM vertex is given by \begin{eqnarray}\label{FEM2}
\hat{\Gamma}^{\mu}(P+Q,P)=\left(\begin{array}{cc} \gamma^{\mu}(P+Q,P) & \Pi^{\mu}(Q)
\\ \bar{\Pi}^{\mu}(Q) & -\gamma^{\mu}(-P,-P-Q)\end{array}\right). \end{eqnarray} This
type of expression was earlier obtained in Ref.\cite{Arseev}. Substituting
Eqs.(\ref{FEM2}) and (\ref{Green}) into the expression (\ref{Kmn1}), we have
\begin{eqnarray}
K^{\mu\nu}(Q)&=&\frac{n}{m}h^{\mu\nu}+\sum_P\big[\gamma^{\mu}(P+Q,P)G(P+Q)\gamma^{\nu}(P,P+Q)G(P)+\Pi^{\mu}(Q)F(P+Q)\gamma^{\nu}(P,P+Q)G(P)\nonumber\\
&-&\gamma^{\mu}(P+Q,P)F(P+Q)\gamma^{\nu}(-P-Q,-P)F(P)+\Pi^{\mu}(Q)G(-P-Q)\gamma^{\nu}(-P-Q,-P)F(P)\nonumber\\
&+&\bar{\Pi}^{\mu}(Q)G(P+Q)\gamma^{\nu}(P,P+Q)F(P)-\gamma^{\mu}(-P,-P-Q)F(P+Q)\gamma^{\nu}(P,P+Q)F(P)\nonumber\\
&+&\bar{\Pi}^{\mu}(Q)F(P+Q)\gamma^{\nu}(-P-Q,-P)G(-P)+\gamma^{\mu}(-P,-P-Q)G(-P-Q)\gamma^{\nu}(-P-Q,-P)G(-P)\big].
\end{eqnarray} Changing variables by $-P\rightarrow P+Q$ for the terms containing
$\gamma_{\nu}(-P-Q,-P)$ and using $F(P)=F(-P)$, we get \begin{eqnarray}\label{Pmunu1}
K^{\mu\nu}(Q)&=&2\sum_P\big[\gamma^{\mu}(P+Q,P)G(P+Q)\gamma^{\nu}(P,P+Q)G(P)+\Pi^{\mu}(Q)F(P+Q)\gamma^{\nu}(P,P+Q)G(P)\nonumber\\
&+&\bar{\Pi}^{\mu}(Q)G(P+Q)\gamma^{\nu}(P,P+Q)F(P)-\gamma^{\mu}(-P,-P-Q)F(P+Q)\gamma^{\nu}(P,P+Q)F(P)\big]+\frac{n}{m}h^{\mu\nu}.
\end{eqnarray} Substituting the expression $F(P)=\Delta G_0(-P)G(P)$ into
Eq.~(\ref{Pmunu1}) and comparing with Eq.~(\ref{KmnO}), one can find the full EM
vertex \begin{eqnarray}\label{Gamma1}
\Gamma^{\mu}(P+Q,P)&=&\gamma^{\mu}(P+Q,P)+\Delta\Pi^{\mu}(Q)G_0(-P-Q)+\Delta\bar{\Pi}^{\mu}(Q)G_0(-P)\nonumber\\
& &-\Delta^2G_0(-P)\gamma^{\mu}(-P,-P-Q)G_0(-P-Q). \end{eqnarray}
The second and
third terms can be shown to contain collective-mode effects \cite{HaoPRL10}. The fourth term corresponds to the Maki-Thompson diagram \cite{OurAnnPhys}.
One can verify that this full interacting vertex obeys the WI (\ref{IWI}) in the
one-dimensional space. Contracting both sides of Eq.(\ref{Gamma1}) with $q^{\mu}$, we
have \begin{eqnarray}
q_{\mu}\Gamma^{\mu}(P+Q,P)&=&G_0^{-1}(P+Q)-G_0^{-1}(P)-2\Sigma(P+Q)+2\Sigma(P)-\frac{\Sigma(P+Q)\Sigma(P)}{\Delta^2}\big(G_0^{-1}(-P)-G_0^{-1}(-P-Q)\big)\nonumber\\
&=&G_0^{-1}(P+Q)-G_0^{-1}(P)-\Sigma(P+Q)+\Sigma(P)=G^{-1}(P+Q)-G^{-1}(P), \end{eqnarray} where we have used the fact
$\Sigma(P)=-\Delta^2G_0(-P)$ for BCS superfluids. Moreover, the $f$-sum rule can be shown to
 be satisfied.

From the expression (\ref{Gamma1}) we can find that the many-particle effects are
indeed included consistently in the correction of EM vertex. The collective modes, which correspond to the poles in the response functions, are  many-particle effects.There is a gapless mode associated with the Nambu-Goldstone mode due to
the spontaneous breaking of the U(1)$_{\textrm{EM}}$ symmetry \cite{HaoPRL10}. Because the
contributions of the Nambu-Goldstone modes are properly included, gauge invariance of our theory is restored.

Now we verify that the $0$-th component of the full EM vertex satisfies the $Q$-limit WI
(\ref{QWI}), from which the compressibility sum rule can be derived \cite{Maebashi09}.
By using $G^{-1}(P)=G^{-1}_0(P)-\Sigma(P)$, this can be shown straightforwardly.
\begin{eqnarray}\label{CSR}
\frac{\partial n}{\partial \mu}&=&2\sum_P\frac{\partial G(P)}{\partial \mu}=-2\sum_PG^2(P)\Big(1-\frac{\partial \Sigma(P)}{\partial \mu}\Big)\nonumber\\
&=&-2\sum_P\Gamma^0(P,P)G(P)\gamma^0(P,P)G(P)=-K^{00}(\omega=0,\mathbf{q}\rightarrow\mathbf{0}).
\end{eqnarray}
For non-interacting Fermi gases, we have $\gamma^0(P+Q,P)=1$ and $\Sigma=0$, hence the $Q$-limit WI is automatically satisfied.

Note that
$\lim_{\mathbf{q}\rightarrow\mathbf{0}}G_0(P+Q)|_{\omega=0}=G_0(P)$, we evaluate
$\Gamma^0(P+Q,P)$ in the limit $\omega=0$ and $\mathbf{q}\rightarrow\mathbf{0}$
\begin{eqnarray}\label{LHS}
\lim_{\mathbf{q}\rightarrow\mathbf{0}}\Gamma^0(P+Q,P)|_{\omega=0}&=&1+\Delta\lim_{\mathbf{q}\rightarrow\mathbf{0}}\big(\Pi^{0}(Q)+\bar{\Pi}^{\mu}(Q)\big)|_{\omega=0}G_0(-P)-\Delta^2G^2_0(-P),\nonumber\\
&=&1-2\Delta\lim_{\mathbf{q}\rightarrow\mathbf{0}}\Pi^{0}_1(Q)|_{\omega=0}G_0(-P)-\Delta^2G^2_0(-P).
\end{eqnarray}
Using $\Sigma(P)=-\Delta^2G_0(-P)$, the RHS of Eq.~(\ref{QWI})  
is
\begin{eqnarray}\label{RHS} 1-\frac{\partial \Sigma(P)}{\partial
\mu}=1+2\Delta\frac{\partial \Delta}{\partial \mu}G_0(-P)-\Delta^2G^2_0(-P),
\end{eqnarray} where the identity
$\partial_{\mu}G_0(-P)=-G^2_0(-P)\partial_{\mu}G^{-1}_0(-P)=-G^2_0(-P)$ has been
applied. Comparing Eqs.~(\ref{LHS}) and (\ref{RHS}), we found that the
$Q$-limit Ward identity holds for BCS theory only when
\begin{eqnarray}\label{t1}
\frac{\partial \Delta}{\partial
\mu}=-\lim_{\mathbf{q}\rightarrow\mathbf{0}}\Pi^{0}_1(Q)|_{\omega=0}.
\end{eqnarray}
This has been shown in Eq.~(\ref{pDpm}) so the $Q$-limit WI is also respected, which then guarantees
the compressibility sum rule.

To compare with the results in the Nambu space, we also list our key results of the CFOP
theory in the one-dimensional space
\begin{itemize} \item Response kernel:
\begin{eqnarray}
K^{\mu\nu}(Q)=2\sum_P\Gamma^{\mu}(P+Q,P)G(P+Q)\gamma^{\nu}(P,P+Q)G(P)+\frac{n}{m}h^{\mu\nu},
\end{eqnarray} \item Full EM interacting vertex:
\begin{eqnarray}\Gamma^{\mu}(P+Q,P)&=&\gamma^{\mu}(P+Q,P)+\Delta\Pi^{\mu}(Q)G_0(-P-Q)+\Delta\bar{\Pi}^{\mu}(Q)G_0(-P)\nonumber\\
& &-\Delta^2G_0(-P)\gamma^{\mu}(-P,-P-Q)G_0(-P-Q), \end{eqnarray} \item Ward
identity: \begin{eqnarray} q_{\mu}\Gamma^{\mu}(P+Q,P)=G^{-1}(P+Q)-G^{-1}(P).
\end{eqnarray} \item $Q$-limit Ward identity: \begin{eqnarray}
\Gamma^0(P,P)=1-\frac{\partial \Sigma(P)}{\partial \mu}, \end{eqnarray} \item $f$-sum
rule: \begin{eqnarray}
\int_{-\infty}^{+\infty}d\omega\omega\chi_{\rho\rho}(\omega,\mathbf{q})=n\frac{q^2}{m}.
\end{eqnarray} \end{itemize}
The WI (or GWI) of the CFOP linear response theory for BCS superfluids guarantees that it is gauge invariant. In Appendix~\ref{app:e} the explicit gauge invariance of the CFOP theory is studied from another point of view based on a ``generalized gauge transformation''.

\subsection{Review of Nambu's Linear Response Theory}\label{4C}
Since we have the full EM vertex $\hat{\Gamma}^{\mu}$
from the CFOP theory, it would be helpful to compare it with the results from
Nambu's integral-equation approach \cite{Nambu60}. A parallel discussion for relativistic BCS superfluids can be
found in Ref.~\cite{OurPRD12}. In BCS theory of conventional superconductors, the self energy is
approximated by an integral equation which consists of a ladder approximation for the
electron-phonon interaction. Nambu proposed that the EM vertex should be corrected in
the same way as that for the self energy. Hence the EM interacting vertex follows an
integral equation \begin{eqnarray}\label{IE}
\hat{\Gamma}^{\mu}(P+Q,P)=\hat{\gamma}^{\mu}(P+Q,P)+g\sum_K\sigma_3\hat{G}(K)\hat{\Gamma}^{\mu}(K+Q,K)\hat{G}(K+Q)\sigma_3.
\end{eqnarray}
If a vertex is a solution to this equation, it automatically satisfies the GWI (\ref{GWI5}).
We give our own proof in Appendix~\ref{app:b0}. Ideally,
if we know how to solve this integral equation, we can further calculate the
gauge-invariant response kernel $K^{\mu\nu}$ by Eq.~(\ref{Kmn1}). Unfortunately, very
little is known about the solution. In general, one may expand
the RHS of the equation as a series of $g$ by the iteration method, and then truncate it at some order.
However, this will not produce a gauge-invariant solution. Moreover, the
integral equation (\ref{IE}) is a vector equation while the GWI (\ref{GWI5}) is a
scalar equation so they have different degrees of freedom. This suggests that there should not be a
rigorous one-to-one correspondence between the solutions to the integral equation and
the EM vertex respecting the GWI. As pointed out by Nambu
\cite{Nambu60}, the integral equation is not only consistent with the GWI associated
with the EM vertex but also consist with the GWIs associated with three other
interaction vertices or gauge transformations (as shown in Eq.(4.4) of
Ref.~\cite{Nambu60}).

In fact, what is equivalent to the GWI (\ref{GWI5}) is the contracted
integral equation given by
\begin{eqnarray}\label{IE1}
q_{\mu}\hat{\Gamma}^{\mu}(P+Q,P)=q_{\mu}\hat{\gamma}^{\mu}(P+Q,P)+g\sum_K\sigma_3\hat{G}(K)q_{\mu}\hat{\Gamma}^{\mu}(K+Q,K)\hat{G}(K+Q)\sigma_3.
\end{eqnarray}
The sufficient condition of this proposition is that the contracted
integral equation (\ref{IE1}) can lead to the GWI, which is proven in Appendix~\ref{app:b0}.
The necessary condition of this proposition
is that any EM vertex obeying GWI must also satisfy Eq.~(\ref{IE1}), but not
necessarily the integral equation (\ref{IE}). Importantly, this is pointed out by Nambu
in his seminal paper \cite{Nambu60} and by Schrieffer \cite{Schrieffer_book}. We
briefly outline the proof here. Substituting Eq.~(\ref{GWI5}) into the RHS of
Eq.~(\ref{IE1}), we have \begin{eqnarray} \textrm{RHS}&=
&q_{\mu}\hat{\gamma}^{\mu}(P+Q,P)+g\sum_K\sigma_3\hat{G}(K)\big(\sigma_3\hat{G}^{-1}(K+Q)-\hat{G}^{-1}(K)\sigma_3\big)\hat{G}(K+Q)\sigma_3\nonumber\\
&=&q_{\mu}\hat{\gamma}^{\mu}(P+Q,P)+2i\Delta\sigma_2=q_{\mu}\hat{\Gamma}^{\mu}(K+Q,K)=\textrm{LHS}, \end{eqnarray}
where
Eq.~(\ref{tmp40}) has been applied. Therefore, any gauge invariant EM vertex
$\hat{\Gamma}^{\prime\mu}$ (including the vertex from the CFOP theory) must be a
solution of the contracted integral equation (\ref{IE1}), but not necessarily a
solution of the integral equation (\ref{IE}). Furthermore, $\hat{\Gamma}^{\prime\mu}$
may differ from the vertex $\hat{\Gamma}^{\mu}$ obtained from Eq.~(\ref{IE}) by a
gauge transformation $\hat{\Gamma}^{\prime\mu}=\hat{\Gamma}^{\mu}+\hat{\chi}^{\mu}$,
where $\hat{\chi}^{\mu}$ satisfies the Lorentz equation $q_{\mu}\hat{\chi}^{\mu}=0$.
Such a gauge transformation may not necessarily be
expressed as $\hat{\chi}^{\mu}=\partial^{\mu}\hat{f}$, where $\hat{f}$ is a matrix in the
Nambu space whose elements are harmonic functions. Another example
is given by $\hat{\chi}^{\mu}=\Pi^{\mu}_1(Q)\hat{C}$, where $\hat{C}$ is an arbitrary
constant matrix in the Nambu space with at least one nonzero element. $\Pi^{\mu}_1(Q)$ is
given by Eqs.~(\ref{tmp4}) and $q_{\mu}\Pi^{\mu}_1(Q)=0$, hence $\hat{\chi}^{\mu}$
satisfies the Lorentz equation.
Under the gauge transformation $\hat{\Gamma}^{\mu}\rightarrow
\hat{\Gamma}^{\mu}+\hat{\chi}^{\mu}$, the $0$-th component of the vertex and the density response function from the CFOP theory transform as \begin{eqnarray}
\hat{\Gamma}^0(P,P)\rightarrow\hat{\Gamma}^0(P,P)-\frac{\partial \Delta}{\partial
\mu}\hat{C},\qquad
\chi_{\rho\rho}(Q)\rightarrow\chi_{\rho\rho}(Q)-\frac{1}{\pi}\textrm{Im}\textrm{Tr}\sum_P\big(\Pi^0_1(Q)\hat{C}\hat{G}(P+Q)\sigma_3\hat{G}(P)\big).
\end{eqnarray}
Then one can verify that the $Q$-limit GWI (\ref{QGWI}) and the $f$-sum
rule (\ref{fsN}) can not be satisfied simultaneously under the above transformation even though the GWI
(\ref{GWI}) is respected anyway.

Here we have two remarks on the consistency of a linear response theory for BCS
superfluids.
Firstly, from the proof given in Appendix~\ref{app:b0}, Nambu's approach tells us that if the self-energy of an interacting Fermi gas
satisfies Eq.(\ref{tmp40}), then the vertex given by the integral equation (\ref{IE}) must satisfies the GWI. However, the theory of
BCS superfluids is not the only theory that satisfies Eq.~(\ref{tmp40}) and as shown in Ref.~\cite{Nambu60} there are theories with other types of symmetries that satisfy Eq.~(\ref{tmp40}). Hence Nambu's vertex given by Eq.~(\ref{IE}) may not be specific to
BCS superfluids and it can be more general.
Secondly, the $Q$-limit GWI and $f$-sum rule are not directly linked to the GWI.
They should be imposed as separate constraints for a consistent linear response
theory. The case of non-interacting Fermi gases is special because the bare EM
vertex $\gamma^{\mu}(P+Q,P)$ is already the full vertex and no correction is needed.
If one formulates a linear response theory and finds a gauge-invariant vertex which obeys the
GWI, one can further calculate the response kernel by Eq.~(\ref{Kmn3}). However, this
response kernel may not satisfy the compressibility sum rule and $f$-sum rule. Therefore,
as we emphasized, the GWI, $Q$-limit GWI, and $f$-sum
rule should be independent constraints for a consistent linear response theory for BCS
superfluids. We cannot fully check whether Nambu's integral-equation approach satisfies all these
criteria because finding an exact solution to the integral equation is the bottleneck,
but the CFOP approach does satisfy all those constraints as we have demonstrated.

\subsection{Comparisons between different linear response theories for BCS
superfluids and Meissner Effect}\label{CS6}
Now we compare the two vertices given by the CFOP theory and
Nambu's integral-equation method. In Ref.\cite{Nambu60}, Nambu managed to solve Eq.~(\ref{IE}) with the aid of
Eq.~(\ref{GWI5}) under the following conditions: (1) only the zeroth order of $g$ is considered, (2) $\omega$ and $q$ are both small, and (3) it is at zero
temperature. Here we compare the result (\ref{FEM}) with Nambu's under the same
condition. Taking the same limits, the response functions $\mathbf{Q}^i_{31}$, $Q_{12}$ and
$Q^0_{13}$ vanish because of the particle-hole symmetry. Therefore
$\Pi^0(Q)=-\bar{\Pi}^0(Q)=-i\frac{Q^0_{23}(Q)}{\tilde{Q}_{22}(Q)}$ and
$\mathbf{\Pi}(Q)=-\bar{\mathbf{\Pi}}(Q)=-i\frac{\mathbf{Q}_{23}(Q)}{\tilde{Q}_{22}(Q)}$.
From the expressions of the response functions, one can verify that  (see
Appendix~\ref{app:b})
\begin{eqnarray}
Q^0_{23}(Q)\simeq-i\frac{N(0)\omega}{2\Delta},\quad\mathbf{Q}_{23}(Q)=-i\frac{N(0)c^2_s}{\Delta}\mathbf{q},\quad\tilde{Q}_{22}(Q)\simeq
-\frac{N(0)}{2\Delta^2}\big(\omega^2-c^2_sq^2\big), \end{eqnarray}
where $N(0)$ is
the density of states at the Fermi energy and
$c_s=\frac{1}{\sqrt{3}}\frac{k_F}{m}=\frac{1}{\sqrt{3}}v_F$ is the speed of sound.
Here $k_F$ is the Fermi momentum defined by $n=\frac{k^3_F}{3\pi^2}$. Therefore we
have \begin{eqnarray}
\Pi^0(Q)=-\bar{\Pi}^0(Q)\simeq\frac{\Delta\omega}{\omega^2-c^2_sq^2},\quad
\mathbf{\Pi}(Q)=-\bar{\mathbf{\Pi}}(Q)\simeq\frac{2\Delta
c^2_s}{\omega^2-c^2_sq^2}\mathbf{q}, \end{eqnarray} and
\begin{eqnarray}\label{eqn:Ga}
\hat{\Gamma}^0(P+Q,P)=\sigma_3+2i\sigma_2\frac{\Delta\omega}{\omega^2-c^2_sq^2},\quad\hat{\mathbf{\Gamma}}(P+Q,P)=\frac{\mathbf{p}+\frac{\mathbf{q}}{2}}{m}+2i\sigma_2\frac{\Delta
c^2_s\mathbf{q}}{\omega^2-c^2_sq^2}. \end{eqnarray}
$\omega=c_sq$ is the dispersion
of the gapless collective mode. These results are exactly the same as those found by Nambu
\cite{Nambu60} by taking the same limits.

Since the CFOP vertex and Nambu's vertex both satisfies the GWI,
 they are at most off by a term $\hat{\chi}_{\mu}$ with
$q_{\mu}\hat{\chi}^{\mu}=0$. $\hat{\chi}_{\mu}$ can be expressed by a harmonic matrix
$\hat{f}$ in the Nambu space of the form $\hat{\chi}_{\mu}=\partial_{\mu}\hat{f}$. In momentum
space, it is $\hat{\chi}_{\mu}=iq_{\mu}\hat{f}$, which vanishes as
$q_{\mu}\rightarrow 0$. Therefore, at zero temperature and in the low frequency and momentum limit, the CFOP
linear response theory agrees with Nambu's approach but in general they can be different.

Before closing our discussions on the density channel, we remark that the collective modes do not contribute to the Meissner effect, where one can show that a finite superfluid density leads to perfect diamagnetism \cite{Walecka}. This remark justifies the standard calculation of the Meissner effect and superfluid density \cite{Walecka}, where one ignores the fluctuations from the order parameter and still obtains the correct results. Although one may use a fully gauge-invariant linear response theory to demonstrate this result \cite{Arseev}, here we directly evaluate the collective-mode contribution in the Meissner effect. Following Ref.~\cite{Walecka}, the Meissner effect is associated with the current-current correlation functions, which can be inferred from the  transverse components of the response kernel. Here we briefly sketch why the collective-mode effects do not contribute to the transverse components of $\tensor{K}^{ij}(0,\mathbf{q})$ as $\mathbf{q}\rightarrow\mathbf{0}$. From Eq.~(\ref{dK}), the response kernel $\tensor{K}^{ij}$ is given by
\begin{eqnarray}\label{M1}
 \tensor{K}^{ij}=\tensor{\tilde{Q}}^{ij}_{33}-\frac{\tilde{Q}_{11}\mathbf{Q}^i_{32}\mathbf{Q}^j_{23}+\tilde{Q}_{22}\mathbf{Q}^i_{31}\mathbf{Q}^j_{13}-2Q_{12}\mathbf{Q}^i_{31}\mathbf{Q}^j_{23}}{\tilde{Q}_{11}\tilde{Q}_{22}-Q_{12}Q_{21}},
\end{eqnarray}
where the second term is associated with the collective modes. A tensor $\tensor{P}^{ij}$ can be decomposed into the longitudinal and the transverse parts $P_{L}$ and $P_T$, where $P_{L}=\hat{\mathbf{q}}\cdot\tensor{P}\cdot\hat{\mathbf{q}}$, $P_T=(\sum_i\tensor{P}^{ii}-P_L)/2$, and $\hat{\mathbf{q}}$ is the unit vector along $\mathbf{q}$. Assuming that $\mathbf{q}$ is parallel to the $z$-axis, in the limit $\mathbf{q}\rightarrow\mathbf{0}$ one can show that $\mathbf{Q}^z_{31}$ and $\mathbf{Q}^z_{32}$ start with linear dependence on $q$. Therefore $\lim_{\mathbf{q}\rightarrow\mathbf{0}}\mathbf{Q}_{3i}\cdot\mathbf{Q}_{3j}=\lim_{\mathbf{q}\rightarrow\mathbf{0}}\hat{\mathbf{q}}\cdot\mathbf{Q}_{3i}\mathbf{Q}_{3j}\cdot \hat{\mathbf{q}}$.
When one evaluates the transverse components of Eq.~(\ref{M1}), all of the collective-mode terms cancel in the limit $\mathbf{q}\rightarrow\mathbf{0}$ so the demonstration of the Meissner effect is insensitive to the collective modes.

\section{Spin Linear Response Theory of
BCS Superfluids}\label{C5}
We now formulate a generalized spin linear
response theory which is similar to its counterpart in the density channel. Using the notation of the
Nambu spinor (\ref{Ns}), the Hamiltonian (\ref{spinH}) with the BCS approximation in
 momentum space is given by \begin{eqnarray}\label{HS0}
H=\sum_{\mathbf{p}}\Psi^{\dagger}_{\mathbf{p}}\xi_{\mathbf{p}}\sigma_3\Psi_{\mathbf{p}}+\sum_{\mathbf{p}\mathbf{q}}
\Psi^{\dagger}_{\mathbf{p}+\mathbf{q}}\big(-\frac{\mathbf{p}+\frac{\mathbf{q}}{2}}{m}\mathbf{A}_{\mathbf{q}}\sigma_3+\Phi_{\mathbf{q}}-\Delta_{\mathbf{q}}\sigma_+-\Delta^*_{-\mathbf{q}}\sigma_-\big)\Psi_{\mathbf{p}}.
\end{eqnarray}
Here the fluctuation of the order parameter $\Delta_{\mathbf{q}}$ is in the spin channel and should not be related to the fluctuation in the density channel.
We follow the same procedure as what we did in the density channel. The order parameter is separated into two parts with one denoting its equilibrium value and the other  denoting the perturbative part.
Furthermore, by introducing the spin
interacting vertex
$\hat{\gamma}^{\mu}_{\textrm{S}}(P+Q,P)\equiv\hat{\gamma}^{\mu}_{\textrm{S}}(\mathbf{p+q},\mathbf{p})=(1,\frac{\mathbf{p}+\frac{\mathbf{q}}{2}}{m}\sigma_3)$
in the Nambu space, the Hamiltonian can be expressed as $H=H_0+H_{\textrm{S}}'$, where
\begin{eqnarray}\label{HS1}
H_0=\sum_{\mathbf{p}}\Psi^{\dagger}_{\mathbf{p}}\hat{E}_{\mathbf{p}}\Psi_{\mathbf{p}},
\quad
H^{\prime}_{\textrm{S}}=\sum_{\mathbf{p}\mathbf{q}}\Psi^{\dagger}_{\mathbf{p}+\mathbf{q}}\big(\Delta_{1\mathbf{q}}\sigma_1+\Delta_{2\mathbf{q}}\sigma_2+A_{\mu\mathbf{q}}\hat{\gamma}_{\textrm{S}}^{\mu}(\mathbf{p}+\mathbf{q},\mathbf{p})\big)\Psi_{\mathbf{p}},
\end{eqnarray}
Similar to its density counterpart, the interacting Hamiltonian can also be cast into the form
\begin{eqnarray}\label{HS2}
H'_{\textrm{S}}=\sum_{\mathbf{p}\mathbf{q}}\Psi^{\dagger}_{\mathbf{p}+\mathbf{q}}\hat{\Phi}^T_{\mathbf{q}}\cdot\hat{\Sigma}_{\textrm{S}}(\mathbf{p}+\mathbf{q},\mathbf{p})\Psi_{\mathbf{p}}
\end{eqnarray}
by introducing the generalized driving potential and generalized interacting vertex
 \begin{eqnarray}\label{GGPS}
\hat{\mathbf{\Phi}}_{\mathbf{q}}=\big(\Delta_{1\mathbf{q}},\Delta_{2\mathbf{q}},A_{\mu\mathbf{q}}\big)^T,\quad
\hat{\mathbf{\Sigma}}_{\textrm{S}}(\mathbf{p}+\mathbf{q},\mathbf{p})=\big(\sigma_1,\sigma_2,\hat{\gamma}^{\mu}_{\textrm{S}}(\mathbf{p}+\mathbf{q},\mathbf{p})\big)^T,
\end{eqnarray}
The Heisenberg operator is defined as
$\mathcal{O}(\tau)=e^{H\tau}\mathcal{O}e^{-H\tau}$. Hence, when the external field is weak, the generalized perturbed current in the
spin linear response theory is given by
\begin{eqnarray}\label{RFS0}
\delta \vec{J}_{\textrm{S}}(\tau,\mathbf{q})=\sum_{\mathbf{p}}\langle\Psi^{\dagger}_{\mathbf{p}}(\tau)\hat{\mathbf{\Sigma}}_{\textrm{S}}(\mathbf{p}+\mathbf{q},\mathbf{p})\Psi_{\mathbf{p}+\mathbf{q}}(\tau)\rangle+\frac{n}{m}\delta^{i3}h^{\mu\nu}A_{\nu}(\tau,\mathbf{q}).
\end{eqnarray}
Here $\delta J^{\mu}_{\textrm{S}3}$ denotes the perturbed spin
current and $\delta J^{\mu}_{\textrm{S}1,2}$ denote the perturbations of the gap function.
The spin linear
response theory can also be written in a matrix form
\begin{eqnarray}
\delta \vec{J}_{\textrm{S}}(\omega,\mathbf{q})&=&\tensor{Q}_{\textrm{S}}(\omega,\mathbf{q})\cdot\hat{\mathbf{\Phi}}(\omega,\mathbf{q})
\nonumber \\ &=&\left( \begin{array}{ccc} Q_{\textrm{S}11}(\omega,\mathbf{q}) &
Q_{\textrm{S}12}(\omega,\mathbf{q}) & Q^{\nu}_{\textrm{S}13}(\omega,\mathbf{q}) \\
Q_{\textrm{S}21}(\omega,\mathbf{q}) & Q_{\textrm{S}22}(\omega,\mathbf{q}) &
Q^{\nu}_{\textrm{S}23}(\omega,\mathbf{q}) \\ Q^{\mu}_{\textrm{S}31}(\omega,\mathbf{q}) &
Q^{\mu}_{\textrm{S}32}(\omega,\mathbf{q}) &
Q^{\mu\nu}_{\textrm{S}33}(\omega,\mathbf{q})+\frac{n}{m}h^{\mu\nu} \end{array}\right)\left(
\begin{array}{ccc} \Delta_{1}(\omega,\mathbf{q}) \\ \Delta_{2}(\omega,\mathbf{q}) \\A_{\nu}(\omega,\mathbf{q})
\end{array}\right). \end{eqnarray}
Following the same steps as what we did previously, the spin response functions are also expressed by
\begin{eqnarray}\label{RFS} Q_{\textrm{S}ij}(\omega,
\mathbf{q})=\textrm{Tr}T\sum_{i\omega_n}\sum_{\mathbf{p}}
\big(\hat{\Sigma}_{\textrm{S}i}(P+Q,P)\hat{G}(P+Q)\hat{\Sigma}_{\textrm{S}j}(P,P+Q)\hat{G}(P)\big),
\end{eqnarray}
The expressions of these response kernels are given in
Appendix~\ref{app:a0}.
One can verify that $Q_{\textrm{S}11}=Q_{11}$, $Q_{\textrm{S}12}=Q_{12}$, $Q_{\textrm{S}21}=Q_{21}$ and $Q_{\textrm{S}22}=Q_{22}$ so the block of the matrix $Q_{\textrm{S}ij}$ for the fluctuations of the order parameter is exactly the same as that in the density channel. It can also be shown that $Q^{\mu}_{\textrm{S}13}=Q^{\mu}_{\textrm{S}23}=Q^{\mu}_{\textrm{S}31}=Q^{\mu}_{\textrm{S}32}=0$.
Hence the perturbation of the order parameter decouples from the perturbation of the spin current. By applying the self-consistent condition $\delta J_{\textrm{S}1,2}=-\frac{2}{g}\Delta_{1,2}$, we obtain
\begin{eqnarray}\label{RFSE}
\left(
\begin{array}{ccc} 0 \\ 0 \\ \delta J^{\mu}_{\textrm{S}3}(\omega,\mathbf{q})
\end{array}\right)=\left( \begin{array}{ccc} \tilde{Q}_{\textrm{S}11}(\omega,\mathbf{q}) &
Q_{\textrm{S}12}(\omega,\mathbf{q}) & 0\\
Q_{\textrm{S}21}(\omega,\mathbf{q}) & \tilde{Q}_{\textrm{S}22}(\omega,\mathbf{q}) &
0 \\ 0 &
0 &
\tilde{Q}^{\mu\nu}_{\textrm{S}}(\omega,\mathbf{q}) \end{array}\right)\left(
\begin{array}{ccc} \Delta_{1}(\omega,\mathbf{q}) \\ \Delta_{2}(\omega,\mathbf{q}) \\A_{\nu}(\omega,\mathbf{q})
\end{array}\right), \end{eqnarray}
where $\tilde{Q}_{\textrm{S}11}=\frac{2}{g}+Q_{\textrm{S}11}$, $\tilde{Q}_{\textrm{S}22}=\frac{2}{g}+Q_{\textrm{S}22}$ and $\tilde{Q}^{\mu\nu}_{\textrm{S}}=Q^{\mu\nu}_{\textrm{S}33}+\frac{n}{m}h^{\mu\nu}$.
The perturbations $\Delta_1$ and $\Delta_2$ in the spin channel can be further shown to be zero, which are very different from their counterparts in the density channel. The spin response function becomes
\begin{eqnarray}
 \delta J^{\mu}_{\textrm{S}3}(\omega,\mathbf{q})=\tilde{Q}^{\mu\nu}_{\textrm{S}}(\omega,\mathbf{q})A_{\nu}(\omega,\mathbf{q}).
 \end{eqnarray}
We emphasize that the fluctuations of the order parameter automatically decouple from the spin response and there is no contribution from the collective modes in the spin response function. The mechanism behind this decoupling is that the BCS order parameter does not break the U(1)$_z$ symmetry. Even though one  treats the spin linear response in the same way as what we did for the EM response, the unbroken U(1)$_z$ symmetry leads to a significantly different result.

The U(1)$_z$ invariance of the linear response theory is satisfied
by the GWI
\begin{equation}\label{GWIS}
q_{\mu}\tilde{Q}_{\textrm{S}}^{\mu\nu}(Q)=0,
\end{equation}
which leads to the
conservation of the spin current $q_{\mu} \delta J_{\textrm{S}}^{\mu}=0$. Before proving this
statement, it is important to notice that the spin interacting vertex obeys the GWI associated
with the U(1)$_z$ symmetry
\begin{eqnarray}\label{WIS}
q_{\mu}\hat{\gamma}^{\mu}_{\textrm{S}}(P+Q,P)=\hat{G}^{-1}_0(P+Q)-\hat{G}^{-1}_0(P)=\hat{G}^{-1}(P+Q)-\hat{G}^{-1}(P).
\end{eqnarray} This leads to the GWI \eqref{GWIS}.
The second equation is due to the fact that the BCS self energy in the Nambu space is give by $\hat{\Sigma}=-\Delta\sigma_1$ and is independent of the four-momentum. The proof of the conservation of the spin current is similar to the derivation shown in Eq.(\ref{WI3}).

The spin susceptibility can be evaluated from the spin response kernel by
$\chi_{\textrm{SS}}=-\frac{1}{\pi}\textrm{Im}Q^{00}_{\textrm{S}33}$. The $f$-sum rule
can be shown to be satisfied in the spin channel. \begin{eqnarray}\label{fsS}
\int_{-\infty}^{+\infty}d\omega\omega\chi_{\textrm{SS}}(\omega,\mathbf{q})=n\frac{q^2}{m}.
\end{eqnarray}
Following the same argument in the density channel, we have
\begin{eqnarray}\label{fsS2}
-\frac{1}{\pi}\int_{-\infty}^{+\infty}d\omega\omega\textrm{Im}Q^{00}_{\textrm{S}33}=-\frac{1}{\pi}\int_{-\infty}^{+\infty}d\omega\textrm{Im}\big(\omega
Q^{00}_{\textrm{S}33})\nonumber =-\frac{1}{\pi}\int_{-\infty}^{+\infty}d\omega
\textrm{Im}\big( \mathbf{q}\cdot \mathbf{Q}^0_{\textrm{S}33}\big). \end{eqnarray}
Comparing Eqs.(\ref{S-Q0i}) to (\ref{Q3i}), we see that
$\mathbf{Q}^{0i}_{\textrm{S}33}=\mathbf{Q}^{0i}_{33}$.
Hence by using the lemma
(\ref{fsl1}) we get \begin{eqnarray}\label{fsS3}
-\frac{1}{\pi}\int_{-\infty}^{+\infty}d\omega\omega\textrm{Im}Q^{00}_{\textrm{S}33}=-\frac{1}{\pi}\int_{-\infty}^{+\infty}d\omega
\textrm{Im}\big( \mathbf{q}\cdot \mathbf{Q}^0_{33}\big)=\frac{nq^2}{m}.
\end{eqnarray}

In the one-dimensional space we follow the same steps in the density channel to
find the U(1)$_z$ gauge-invariant spin vertex $\Gamma^{\mu}_{\textrm{S}\sigma}$ which
satisfies the WI (\ref{SWI}). The spin response kernel can be expressed by
\begin{eqnarray}\label{RFSF1}
\tilde{Q}^{\mu\nu}_{\textrm{S}}(Q)=\sum_{P}\sum_{\sigma}
\Gamma^{\mu}_{\textrm{S}\sigma}(P+Q,P)G(P+Q)\gamma^{\nu}_{\textrm{S}\sigma}(P,P+Q)G(P)+\frac{n}{m}h^{\mu\nu},
\end{eqnarray}
where the bare spin interacting vertex
$\gamma^{\mu}_{\textrm{S}\sigma}$ is given in Sec.~\ref{C2}. Importantly, Eq.\eqref{RFSF1} should give the same expression for the spin response function as the $33$-component of Eq.~(\ref{RFS}) when the $h^{\mu\nu}$ term is included
\begin{eqnarray}\label{RFS1}
\tilde{Q}^{\mu\nu}_{\textrm{S}}(Q)=\textrm{Tr}\sum_{P}
\big(\hat{\gamma}^{\mu}_{\textrm{S}}(P+Q,P)\hat{G}(P+Q)\hat{\gamma}^{\nu}_{\textrm{S}}(P,P+Q)\hat{G}(P)\big)+\frac{n}{m}h^{\mu\nu}.
\end{eqnarray}
Note that the spin interacting vertex in the Nambu space can be expressed as
\begin{eqnarray}\label{GSN}
\hat{\gamma}^{\mu}_{\textrm{S}}(P+Q,P)=\left(\begin{array}{cc}\gamma^{\mu}_{\textrm{S}\uparrow}(P+Q,P)
& 0\\0 & -\gamma^{\mu}_{\textrm{S}\downarrow}(-P,-P-Q)\end{array}\right).
\end{eqnarray}
Substituting Eq.~(\ref{GSN}) and the propagator (\ref{NGP}) into the
expression (\ref{RFS1}), we get
\begin{eqnarray}\label{RFSF2} &
&\tilde{Q}^{\mu\nu}_{\textrm{S}}(Q) \nonumber\\
&=&\frac{n}{m}h^{\mu\nu}+\sum_{P}\big(\gamma^{\mu}_{\textrm{S}\uparrow}(P+Q,P)G(P+Q)\gamma^{\nu}_{\textrm{S}\uparrow}(P,P+Q)G(P)-\gamma^{\mu}_{\textrm{S}\uparrow}(P+Q,P)F(P+Q)\gamma^{\nu}_{\textrm{S}\downarrow}(-P-Q,-P)F(P)
\nonumber\\ &
-&\gamma^{\mu}_{\textrm{S}\downarrow}(-P,-P-Q)F(P+Q)\gamma^{\nu}_{\textrm{S}\uparrow}(P,P+Q)F(P)+\gamma^{\mu}_{\textrm{S}\downarrow}(-P,-P-Q)G(-P-Q)\gamma^{\nu}_{\textrm{S}\downarrow}(-P-Q,-P)G(-P)\big)\nonumber\\
&=&\frac{n}{m}h^{\mu\nu}+\sum_{P}\big(\gamma^{\mu}_{\textrm{S}\uparrow}(P+Q,P)G(P+Q)\gamma^{\nu}_{\textrm{S}\uparrow}(P,P+Q)G(P)-\gamma^{\mu}_{\textrm{S}\uparrow}(-P,-P-Q)F(P)\gamma^{\nu}_{\textrm{S}\downarrow}(P,P+Q)F(P+Q)
\nonumber\\
&-&\gamma^{\mu}_{\textrm{S}\downarrow}(-P,-P-Q)F(P+Q)\gamma^{\nu}_{\textrm{S}\uparrow}(P,P+Q)F(P)+\gamma^{\mu}_{\textrm{S}\downarrow}(P+Q,P)G(P)\gamma^{\nu}_{\textrm{S}\downarrow}(P,P+Q)G(P+Q)\big),
\end{eqnarray}
where we have changed variables by $-P\rightarrow P+Q$ for the terms
containing $\gamma^{\nu}_{\textrm{S}\downarrow}(-P-Q,-P)$ and applied $F(P)=F(-P)$.
Using $F(P)=\Delta G_0(-P)G(P)$ again and Eq.(\ref{RFSF1}),
one can see that the following expression guarantees that the expressions for $\tilde{Q}^{\mu\nu}_{\textrm{S}}(Q)$ are the same.
\begin{eqnarray}\label{SV0}
\Gamma^{\mu}_{\textrm{S}\sigma}(P+Q,P)=\gamma^{\mu}_{\textrm{S}\sigma}(P+Q,P)-\Delta^2G_0(-P)\gamma^{\mu}_{\textrm{S}\bar{\sigma}}(-P,-P-Q)G_0(-P-Q)
\end{eqnarray} for $\sigma=\uparrow,\downarrow$.
The second term corresponds to the
Maki-Thompson diagram \cite{OurAnnPhys}.
Now we show that this full spin interacting vertex satisfies the Ward identity
(\ref{SWI}). Using $S_{\sigma}=-S_{\bar{\sigma}}$ and Eq.(\ref{BWI}), we have
\begin{eqnarray}\label{SWI}
q_{\mu}\Gamma^{\mu}_{\textrm{S}\sigma}(P+Q,P)&=&S_{\sigma}\big(G^{-1}_{0}(P+Q)-G^{-1}_{0}(P)\big)+S_{\sigma}\Delta^2G_0(-P)G_0(-P-Q)\big(G^{-1}_{0}(-P)-G^{-1}_{0}(-P-Q)\big)\nonumber\\
&=&S_{\sigma}\big(G^{-1}_{0}(P+Q)-G^{-1}_{0}(P)\big)-S_{\sigma}\big(\Sigma(P+Q)-\Sigma(P)\big)
=S_{\sigma}\big(G^{-1}(P+Q)-G^{-1}(P)\big). \end{eqnarray}

Finally we summarize the
central results in this section.
In the Nambu space, we have
\begin{itemize} \item Response kernel: \begin{eqnarray}
\tilde{Q}^{\mu\nu}_{\textrm{S}}(Q)=\textrm{Tr}\sum_{P}
\big(\hat{\gamma}^{\mu}_{\textrm{S}}(P+Q,P)\hat{G}(P+Q)\hat{\gamma}^{\nu}_{\textrm{S}}(P,P+Q)\hat{G}(P)\big)+\frac{n}{m}h^{\mu\nu},
\end{eqnarray} \item Spin interacting vertex:
\begin{eqnarray}\hat{\gamma}^{\mu}_{\textrm{S}}(P+Q,P)=(1,\frac{\mathbf{p}+\frac{\mathbf{q}}{2
}}{m}\sigma_3),
\end{eqnarray} \item Generalized Ward identity: \begin{eqnarray}
q_{\mu}\hat{\gamma}^{\mu}_{\textrm{S}}(P+Q,P)=\hat{G}^{-1}(P+Q)-\hat{G}^{-1}(P).
\end{eqnarray} \end{itemize}

In the one-dimensional space, we have
\begin{itemize} \item Response kernel:
\begin{eqnarray} \tilde{Q}^{\mu\nu}_{\textrm{S}}(Q)=\sum_{P}\sum_{\sigma}
\Gamma^{\mu}_{\textrm{S}\sigma}(P+Q,P)G(P+Q)\gamma^{\nu}_{\textrm{S}\sigma}(P,P+Q)G(P)+\frac{n}{m}h^{\mu\nu},
\end{eqnarray} \item Full spin interacting vertex:
\begin{eqnarray}\Gamma^{\mu}_{\textrm{S}\sigma}(P+Q,P)=\gamma^{\mu}_{\textrm{S}\sigma}(P+Q,P)+\Delta^2G_0(-P)\gamma^{\mu}_{\textrm{S}\bar{\sigma}}(-P,-P-Q)G_0(-P-Q),
\end{eqnarray} \item Ward identity: \begin{eqnarray}
q_{\mu}\Gamma^{\mu}_{\textrm{S}\sigma}(P+Q,P)=S_{\sigma}\big(G^{-1}(P+Q)-G^{-1}(P)\big).
\end{eqnarray} \end{itemize}

In both spaces, the $f$-sum rules are satisfied \begin{eqnarray}
\int_{-\infty}^{+\infty}d\omega\omega\chi_{\textrm{SS}}(\omega,\mathbf{q})=n\frac{q^2}{m}.
\end{eqnarray}

Below $T_c$, the
U(1)$_{\textrm{EM}}$ symmetry is broken by the condensed Cooper pairs while the U(1)$_z$
symmetry remains intact. Hence the two linear response theories produce significantly
different results below $T_c$. Importantly, the collective modes coming from the breaking U(1)$_{\textrm{EM}}$ symmetry
only couples to the density response function but decouples from the spin response function.
This indicates
that \textit{the difference between the density and spin channels arises only in the
presence of a condensate that only breaks the} $U(1)_{\textrm{EM}}$ \textit{symmetry}.
Above $T_c$, both U(1) symmetries are respected since there is no
Cooper-pair condensation. The effective Lagrangian after the BCS approximation is identical
to that of a non-interacting Fermi gas when the gap $\Delta$ vanishes. By dropping
$Q_{1i}$ and $Q_{2i}$ which are not defined above $T_c$, one can verify that the
response functions of the two linear response theories give the same result above $T_c$.
In the presence of pairing fluctuation effects, the
amplitude of the pairs are not necessarily the order parameter since finite-momentum
pairs may coexist with the condensate of Cooper pairs \cite{OurReview}. The difference between
the two response functions could be shown to be still valid and one can use the
difference between the density and spin structure factors to construct a quantity
similar to an order parameter for detecting the phase coherence of atomic Fermi gases
\cite{HaoPRL10}.

\section{Conclusion}\label{Conclusion}
We have shown that the CFOP approach of the linear response of BCS superfluid is a computational manageable scheme that satisfies important constraints including Ward identity, $f$ sum rule, $Q$-limit Ward identity, and compressibility sum rule that guarantee charge conservation and a consistent expression for the compressibility. The CFOP formalism provides a paradigm for studying linear response theories in interacting many-body systems in the presence of spontaneous symmetry breaking. The spin linear response theory complements the story of the CFOP theory and demonstrates the different roles played by the collective modes in the superfluid phase. Going beyond mean-field BCS theory requires considerations of non-condensed pairs and there have been different approaches \cite{OurAnnPhys}. We emphasize that linear response theories of those beyond-BCS theories should be subject to the same consistency constraints discussed here. In addition to conventional superconductors \cite{Arseev}, our formalism may be useful in the study of ultra-cold atoms \cite{HaoPRL10,HaoNJP11} and nuclear physics \cite{ReddyPRC04,Gusakov10} where linear response theories of BCS superfluids are frequently implemented.

\section*{Acknowledgement}
We thank Prof. K. Levin for helping prepare this paper. Hao Guo thanks the support by National Natural Science Foundation of China (Grants No. 11204032) and Natural Science Foundation of Jiangsu Province, China (SBK201241926). C. C. C. acknowledges the support of the U.S. Department of Energy through the LANL/LDRD Program.

\appendix

\section{Detailed expressions for response functions}\label{app:a0}
The following are the response functions for non-interacting Fermi gases:
\begin{eqnarray}\label{FS-Q00}
Q_0^{00}(\omega,\mathbf{q})&=&-2\sum_{\mathbf{p}}\frac{\xi^+_{\mathbf{p}}-\xi^-_{\mathbf{p}}}{\omega^2-(\xi^+_{\mathbf{p}}-\xi^-_{\mathbf{p}})^2}[f(\xi^+_{\mathbf{p}})-f(\xi^-_{\mathbf{p}})],
\end{eqnarray}

\begin{eqnarray}\label{FS-Q0i}
\mathbf{Q}_0^{0i}(\omega,\mathbf{q})=\mathbf{Q}_0^{i0}(\omega,\mathbf{q})=-2\omega\sum_{\mathbf{p}}\frac{\mathbf{p}^i}{m}\frac{f(\xi^+_{\mathbf{p}})-f(\xi^-_{\mathbf{p}})}{\omega^2-(\xi^+_{\mathbf{p}}-\xi^-_{\mathbf{p}})^2},
\end{eqnarray}

\begin{eqnarray}\label{FS-Qij}
\tensor{Q}_0^{ij}(\omega,\mathbf{q})&=&-2\sum_{\mathbf{p}}\frac{\mathbf{p}^i\mathbf{p}^j}{m^2}\frac{\xi^+_{\mathbf{p}}-\xi^-_{\mathbf{p}}}{\omega^2-(\xi^+_{\mathbf{p}}-\xi^-_{\mathbf{p}})^2}[f(\xi^+_{\mathbf{p}})-f(\xi^-_{\mathbf{p}})].
\end{eqnarray}

The following are the EM response functions of BCS superfluids from the CFOP theory:
\begin{eqnarray}
Q_{11}(\omega,\mathbf{q})&=&\sum_{\mathbf{p}}\Big[\big(1+\frac{\xi^+_{\mathbf{p}}\xi^-_{\mathbf{p}}-\Delta^2}{E^+_{\mathbf{p}}E^-_{\mathbf{p}}}\big)\frac{E^+_{\mathbf{p}}+E^-_{\mathbf{p}}}{\omega^2-(E^+_{\mathbf{p}}+E^-_{\mathbf{p}})^2}[1-f(E^+_{\mathbf{p}})-f(E^-_{\mathbf{p}})] \nonumber \\
& &\quad-\big(1-\frac{\xi^+_{\mathbf{p}}\xi^-_{\mathbf{p}}-\Delta^2}{E^+_{\mathbf{p}}E^-_{\mathbf{p}}}\big)\frac{E^+_{\mathbf{p}}-E^-_{\mathbf{p}}}{\omega^2-(E^+_{\mathbf{p}}-E^-_{\mathbf{p}})^2}[f(E^+_{\mathbf{p}})-f(E^-_{\mathbf{p}})]\Big],
\end{eqnarray}
\begin{eqnarray}
Q_{12}(\omega,\mathbf{q})=-Q_{21}(\omega,\mathbf{q})=-i\omega\sum_{\mathbf{p}}\Big[\big(\frac{\xi^+_{\mathbf{p}}}{E^+_{\mathbf{p}}}+\frac{\xi^-_{\mathbf{p}}}{E^-_{\mathbf{p}}}\big)\frac{1-f(E^+_{\mathbf{p}})-f(E^-_{\mathbf{p}})}{\omega^2-(E^+_{\mathbf{p}}+E^-_{\mathbf{p}})^2}-\big(\frac{\xi^+_{\mathbf{p}}}{E^+_{\mathbf{p}}}-\frac{\xi^-_{\mathbf{p}}}{E^-_{\mathbf{p}}}\big)\frac{f(E^+_{\mathbf{p}})-f(E^-_{\mathbf{p}})}{\omega^2-(E^+_{\mathbf{p}}-E^-_{\mathbf{p}})^2}\Big],
\end{eqnarray}
\begin{eqnarray}
Q^0_{13}(\omega,\mathbf{q})=Q^0_{31}(\omega,\mathbf{q})=\Delta\sum_{\mathbf{p}}\frac{\xi^+_{\mathbf{p}}+\xi^-_{\mathbf{p}}}{E^+_{\mathbf{p}}E^-_{\mathbf{p}}}\Big[\frac{(E^+_{\mathbf{p}}+E^-_{\mathbf{p}})[1-f(E^+_{\mathbf{p}})-f(E^-_{\mathbf{p}})]}{\omega^2-(E^+_{\mathbf{p}}+E^-_{\mathbf{p}})^2}+\frac{(E^+_{\mathbf{p}}-E^-_{\mathbf{p}})[f(E^+_{\mathbf{p}})-f(E^-_{\mathbf{p}})]}{\omega^2-(E^+_{\mathbf{p}}-E^-_{\mathbf{p}})^2}\Big],
\end{eqnarray}
\begin{eqnarray}
\mathbf{Q}^i_{13}(\omega,\mathbf{q})=\mathbf{Q}_{31}^i(\omega,\mathbf{q})=\sum_{\mathbf{p}}\frac{\mathbf{p}^i}{m}\frac{\Delta\omega}{E^+_{\mathbf{p}}E^-_{\mathbf{p}}} \Big[\frac{(E^+_{\mathbf{p}}-E^-_{\mathbf{p}})[1-f(E^+_{\mathbf{p}})-f(E^-_{\mathbf{p}})]}{\omega^2-(E^+_{\mathbf{p}}+E^-_{\mathbf{p}})^2}+\frac{(E^+_{\mathbf{p}}+E^-_{\mathbf{p}})[f(E^+_{\mathbf{p}})-f(E^-_{\mathbf{p}})]}{\omega^2-(E^+_{\mathbf{p}}-E^-_{\mathbf{p}})^2}\Big],
\end{eqnarray}
\begin{eqnarray}
Q_{22}(\omega,\mathbf{q})&=&\sum_{\mathbf{p}}\Big[\big(1+\frac{\xi^+_{\mathbf{p}}\xi^-_{\mathbf{p}}+\Delta^2}{E^+_{\mathbf{p}}E^-_{\mathbf{p}}}\big)\frac{E^+_{\mathbf{p}}+E^-_{\mathbf{p}}}{\omega^2-(E^+_{\mathbf{p}}+E^-_{\mathbf{p}})^2}[1-f(E^+_{\mathbf{p}})-f(E^-_{\mathbf{p}})] \nonumber \\
& &\quad-\big(1-\frac{\xi^+_{\mathbf{p}}\xi^-_{\mathbf{p}}+\Delta^2}{E^+_{\mathbf{p}}E^-_{\mathbf{p}}}\big)\frac{E^+_{\mathbf{p}}-E^-_{\mathbf{p}}}{\omega^2-(E^+_{\mathbf{p}}-E^-_{\mathbf{p}})^2}[f(E^+_{\mathbf{p}})-f(E^-_{\mathbf{p}})]\Big],
\end{eqnarray}
\begin{eqnarray}
Q^0_{23}(\omega,\mathbf{q})=-Q^0_{32}(\omega,\mathbf{q})=i\sum_{\mathbf{p}}\frac{\Delta\omega}{E^+_{\mathbf{p}}E^-_{\mathbf{p}}}\Big[\frac{(E^+_{\mathbf{p}}+E^-_{\mathbf{p}})[1-f(E^+_{\mathbf{p}})-f(E^-_{\mathbf{p}})]}{\omega^2-(E^+_{\mathbf{p}}+E^-_{\mathbf{p}})^2}+\frac{(E^+_{\mathbf{p}}-E^-_{\mathbf{p}})[f(E^+_{\mathbf{p}})-f(E^-_{\mathbf{p}})]}{\omega^2-(E^+_{\mathbf{p}}-E^-_{\mathbf{p}})^2}\Big],
\end{eqnarray}
\begin{eqnarray}
\mathbf{Q}^i_{23}(\omega,\mathbf{q})=-\mathbf{Q}^i_{32}(\omega,\mathbf{q})=i\Delta\sum_{\mathbf{p}}\frac{\mathbf{p}^i}{m}\frac{\xi^+_{\mathbf{p}}-\xi^-_{\mathbf{p}}}{E^+_{\mathbf{p}}E^-_{\mathbf{p}}} \Big[\frac{(E^+_{\mathbf{p}}+E^-_{\mathbf{p}})[1-f(E^+_{\mathbf{p}})-f(E^-_{\mathbf{p}})]}{\omega^2-(E^+_{\mathbf{p}}+E^-_{\mathbf{p}})^2}+\frac{(E^+_{\mathbf{p}}-E^-_{\mathbf{p}})[f(E^+_{\mathbf{p}})-f(E^-_{\mathbf{p}})]}{\omega^2-(E^+_{\mathbf{p}}-E^-_{\mathbf{p}})^2}\Big],
\end{eqnarray}
\begin{eqnarray}
Q_{33}^{00}(\omega,\mathbf{q})&=&\sum_{\mathbf{p}}\Big[\big(1-\frac{\xi^+_{\mathbf{p}}\xi^-_{\mathbf{p}}-\Delta^2}{E^+_{\mathbf{p}}E^-_{\mathbf{p}}}\big)\frac{E^+_{\mathbf{p}}+E^-_{\mathbf{p}}}{\omega^2-(E^+_{\mathbf{p}}+E^-_{\mathbf{p}})^2}[1-f(E^+_{\mathbf{p}})-f(E^-_{\mathbf{p}})] \nonumber \\
& &\quad-\big(1+\frac{\xi^+_{\mathbf{p}}\xi^-_{\mathbf{p}}-\Delta^2}{E^+_{\mathbf{p}}E^-_{\mathbf{p}}}\big)\frac{E^+_{\mathbf{p}}-E^-_{\mathbf{p}}}{\omega^2-(E^+_{\mathbf{p}}-E^-_{\mathbf{p}})^2}[f(E^+_{\mathbf{p}})-f(E^-_{\mathbf{p}})]\Big].
\end{eqnarray}
\begin{eqnarray}
\tensor{Q}_{33}^{ij}(\omega,\mathbf{q})&=&\sum_{\mathbf{p}}\frac{\mathbf{p}^i\mathbf{p}^j}{m^2}\Big[\big(1-\frac{\xi^+_{\mathbf{p}}\xi^-_{\mathbf{p}}+\Delta^2}{E^+_{\mathbf{p}}E^-_{\mathbf{p}}}\big)\frac{E^+_{\mathbf{p}}+E^-_{\mathbf{p}}}{\omega^2-(E^+_{\mathbf{p}}+E^-_{\mathbf{p}})^2}[1-f(E^+_{\mathbf{p}})-f(E^-_{\mathbf{p}})] \nonumber \\
& &\qquad\quad-\big(1+\frac{\xi^+_{\mathbf{p}}\xi^-_{\mathbf{p}}+\Delta^2}{E^+_{\mathbf{p}}E^-_{\mathbf{p}}}\big)\frac{E^+_{\mathbf{p}}-E^-_{\mathbf{p}}}{\omega^2-(E^+_{\mathbf{p}}-E^-_{\mathbf{p}})^2}[f(E^+_{\mathbf{p}})-f(E^-_{\mathbf{p}})]\Big],
\end{eqnarray}
\begin{eqnarray}\label{Q3i}
\mathbf{Q}^{0i}_{33}(\omega,\mathbf{q})=\mathbf{Q}^{i0}_{33}(\omega,\mathbf{q})=\omega\sum_{\mathbf{p}}\frac{\mathbf{p}^i}{m}\Big[\big(\frac{\xi^+_{\mathbf{p}}}{E^+_{\mathbf{p}}}-\frac{\xi^-_{\mathbf{p}}}{E^-_{\mathbf{p}}}\big)\frac{1-f(E^+_{\mathbf{p}})-f(E^-_{\mathbf{p}})}{\omega^2-(E^+_{\mathbf{p}}+E^-_{\mathbf{p}})^2}-\big(\frac{\xi^+_{\mathbf{p}}}{E^+_{\mathbf{p}}}+\frac{\xi^-_{\mathbf{p}}}{E^-_{\mathbf{p}}}\big)\frac{f(E^+_{\mathbf{p}})-f(E^-_{\mathbf{p}})}{\omega^2-(E^+_{\mathbf{p}}-E^-_{\mathbf{p}})^2}\Big].
\end{eqnarray}

The following are spin response functions of BCS Superfluids following the same structure of the CFOP theory
\begin{eqnarray}
Q_{\textrm{S}11}(\omega,\mathbf{q})&=&\sum_{\mathbf{p}}\Big[\big(1+\frac{\xi^+_{\mathbf{p}}\xi^-_{\mathbf{p}}-\Delta^2}{E^+_{\mathbf{p}}E^-_{\mathbf{p}}}\big)\frac{E^+_{\mathbf{p}}+E^-_{\mathbf{p}}}{\omega^2-(E^+_{\mathbf{p}}+E^-_{\mathbf{p}})^2}[1-f(E^+_{\mathbf{p}})-f(E^-_{\mathbf{p}})] \nonumber \\
& &\quad-\big(1-\frac{\xi^+_{\mathbf{p}}\xi^-_{\mathbf{p}}-\Delta^2}{E^+_{\mathbf{p}}E^-_{\mathbf{p}}}\big)\frac{E^+_{\mathbf{p}}-E^-_{\mathbf{p}}}{\omega^2-(E^+_{\mathbf{p}}-E^-_{\mathbf{p}})^2}[f(E^+_{\mathbf{p}})-f(E^-_{\mathbf{p}})]\Big],
\end{eqnarray}
\begin{eqnarray}
Q_{\textrm{S}12}(\omega,\mathbf{q})=-Q_{\textrm{S}21}(\omega,\mathbf{q})=-i\omega\sum_{\mathbf{p}}\Big[\big(\frac{\xi^+_{\mathbf{p}}}{E^+_{\mathbf{p}}}+\frac{\xi^-_{\mathbf{p}}}{E^-_{\mathbf{p}}}\big)\frac{1-f(E^+_{\mathbf{p}})-f(E^-_{\mathbf{p}})}{\omega^2-(E^+_{\mathbf{p}}+E^-_{\mathbf{p}})^2}-\big(\frac{\xi^+_{\mathbf{p}}}{E^+_{\mathbf{p}}}-\frac{\xi^-_{\mathbf{p}}}{E^-_{\mathbf{p}}}\big)\frac{f(E^+_{\mathbf{p}})-f(E^-_{\mathbf{p}})}{\omega^2-(E^+_{\mathbf{p}}-E^-_{\mathbf{p}})^2}\Big],
\end{eqnarray}
\begin{eqnarray}
Q_{\textrm{S}22}(\omega,\mathbf{q})&=&\sum_{\mathbf{p}}\Big[\big(1+\frac{\xi^+_{\mathbf{p}}\xi^-_{\mathbf{p}}+\Delta^2}{E^+_{\mathbf{p}}E^-_{\mathbf{p}}}\big)\frac{E^+_{\mathbf{p}}+E^-_{\mathbf{p}}}{\omega^2-(E^+_{\mathbf{p}}+E^-_{\mathbf{p}})^2}[1-f(E^+_{\mathbf{p}})-f(E^-_{\mathbf{p}})] \nonumber \\
& &\quad-\big(1-\frac{\xi^+_{\mathbf{p}}\xi^-_{\mathbf{p}}+\Delta^2}{E^+_{\mathbf{p}}E^-_{\mathbf{p}}}\big)\frac{E^+_{\mathbf{p}}-E^-_{\mathbf{p}}}{\omega^2-(E^+_{\mathbf{p}}-E^-_{\mathbf{p}})^2}[f(E^+_{\mathbf{p}})-f(E^-_{\mathbf{p}})]\Big],
\end{eqnarray}
\begin{eqnarray}\label{S-0}
Q^{\mu}_{\textrm{S}13}(\omega,\mathbf{q})=Q^{\mu}_{\textrm{S}23}(\omega,\mathbf{q})=Q^{\mu}_{\textrm{S}31}(\omega,\mathbf{q})=Q^{\mu}_{\textrm{S}32}(\omega,\mathbf{q})=0,
\end{eqnarray}
\begin{eqnarray}\label{S-Q00}
Q_{\textrm{S}33}^{00}(\omega,\mathbf{q})&=&\sum_{\mathbf{p}}\Big\{\big(1-\frac{\xi^+_{\mathbf{p}}\xi^-_{\mathbf{p}}+\Delta^2}{E^+_{\mathbf{p}}E^-_{\mathbf{p}}}\big)\frac{E^+_{\mathbf{p}}+E^-_{\mathbf{p}}}{\omega^2-(E^+_{\mathbf{p}}+E^-_{\mathbf{p}})^2}[1-f(E^+_{\mathbf{p}})-f(E^-_{\mathbf{p}})] \nonumber \\
& &\quad-\big(1+\frac{\xi^+_{\mathbf{p}}\xi^-_{\mathbf{p}}+\Delta^2}{E^+_{\mathbf{p}}E^-_{\mathbf{p}}}\big)\frac{E^+_{\mathbf{p}}-E^-_{\mathbf{p}}}{\omega^2-(E^+_{\mathbf{p}}-E^-_{\mathbf{p}})^2}[f(E^+_{\mathbf{p}})-f(E^-_{\mathbf{p}})]\Big\},
\end{eqnarray}

\begin{eqnarray}\label{S-Q0i}
\mathbf{Q}_{\textrm{S}33}^{0i}(\omega,\mathbf{q})=\mathbf{Q}_{\textrm{S}33}^{i0}(\omega,\mathbf{q})=\omega\sum_{\mathbf{p}}\frac{\mathbf{p}}{m}\Big\{\big(\frac{\xi^+_{\mathbf{p}}}{E^+_{\mathbf{p}}}-\frac{\xi^-_{\mathbf{p}}}{E^-_{\mathbf{p}}}\big)\frac{1-f(E^+_{\mathbf{p}})-f(E^-_{\mathbf{p}})}{\omega^2-(E^+_{\mathbf{p}}+E^-_{\mathbf{p}})^2}-\big(\frac{\xi^+_{\mathbf{p}}}{E^+_{\mathbf{p}}}+\frac{\xi^-_{\mathbf{p}}}{E^-_{\mathbf{p}}}\big)\frac{f(E^+_{\mathbf{p}})-f(E^-_{\mathbf{p}})}{\omega^2-(E^+_{\mathbf{p}}-E^-_{\mathbf{p}})^2}\Big\},
\end{eqnarray}

\begin{eqnarray}\label{S-Qij}
\tensor{Q}_{\textrm{S}33}^{ij}(\omega,\mathbf{q})&=&\sum_{\mathbf{p}}\frac{\mathbf{p}\mathbf{p}}{m^2}\Big\{\big(1-\frac{\xi^+_{\mathbf{p}}\xi^-_{\mathbf{p}}-\Delta^2}{E^+_{\mathbf{p}}E^-_{\mathbf{p}}}\big)\frac{E^+_{\mathbf{p}}+E^-_{\mathbf{p}}}{\omega^2-(E^+_{\mathbf{p}}+E^-_{\mathbf{p}})^2}[1-f(E^+_{\mathbf{p}})-f(E^-_{\mathbf{p}})] \nonumber \\
& &\qquad\quad-\big(1+\frac{\xi^+_{\mathbf{p}}\xi^-_{\mathbf{p}}-\Delta^2}{E^+_{\mathbf{p}}E^-_{\mathbf{p}}}\big)\frac{E^+_{\mathbf{p}}-E^-_{\mathbf{p}}}{\omega^2-(E^+_{\mathbf{p}}-E^-_{\mathbf{p}})^2}[f(E^+_{\mathbf{p}})-f(E^-_{\mathbf{p}})]\Big\}.
\end{eqnarray}

Here we outline the proof of Eq.~(\ref{S-0}). When the index $\mu=0$, from Eq.~(\ref{RFS}) we have
\begin{eqnarray}
Q^{0}_{\textrm{S}13}(\omega,\mathbf{q})
&=&\sum_P\big(G(P)F(P+Q)-G(-P-Q)F(P)+G(P+Q)F(P)-G(-P)F(P)\big).
\end{eqnarray}
Changing variables by $P\rightarrow-P-Q$ and using the fact $F(P)=F(-P)$, the second and fourth terms inside the bracket become $G(P)F(P+Q)$ and $G(P+Q)F(P)$ respectively, which cancel the first and third term inside the bracket respectively. Hence we have $Q^{0}_{\textrm{S}13}(\omega,\mathbf{q})=0$.
Similarly, if the index $\mu=i$, we have
\begin{eqnarray}
\mathbf{Q}_{\textrm{S}13}(\omega,\mathbf{q})&=&\sum_P\frac{\mathbf{p}+\frac{\mathbf{q}}{2}}{m}\big(G(P)F(P+Q)+G(-P-Q)F(P)+G(P+Q)F(P)+G(-P)F(P)\big).
\end{eqnarray}
Here we change variables by $P\rightarrow-P-Q$ again so the pre-factor $\frac{\mathbf{p}+\frac{\mathbf{q}}{2}}{m}\rightarrow-\frac{\mathbf{p}+\frac{\mathbf{q}}{2}}{m}$. Hence all terms inside the bracket cancel out and we conclude that $\mathbf{Q}_{\textrm{S}13}(\omega,\mathbf{q})=0$. Following the same steps, one can prove that $Q^{\mu}_{\textrm{S}23}(\omega,\mathbf{q})=Q^{\mu}_{\textrm{S}31}(\omega,\mathbf{q})=Q^{\mu}_{\textrm{S}32}(\omega,\mathbf{q})=0$.

\section{Proof of the Lemma (\ref{fsl1})}\label{app:a1}
From the expression shown in Appendix~\ref{app:a0}, we have
\begin{eqnarray}\label{fsE1}
\mathbf{Q}^0_{33}(\omega,\mathbf{q})&=&\sum_{\mathbf{p}}\frac{\omega}{2}\frac{\mathbf{p}}{m}\Big\{\big(\frac{\xi^+_{\mathbf{p}}}{E^+_{\mathbf{p}}}-\frac{\xi^-_{\mathbf{p}}}{E^-_{\mathbf{p}}}\big)\frac{1-f(E^+_{\mathbf{p}})-f(E^-_{\mathbf{p}})}{E^+_{\mathbf{p}}+E^-_{\mathbf{p}}}
\Big(\frac{1}{\omega-E^+_{\mathbf{p}}-E^-_{\mathbf{p}}}-\frac{1}{\omega+E^+_{\mathbf{p}}+E^-_{\mathbf{p}}}\Big)\nonumber\\
&
&\qquad\qquad-\big(\frac{\xi^+_{\mathbf{p}}}{E^+_{\mathbf{p}}}+\frac{\xi^-_{\mathbf{p}}}{E^-_{\mathbf{p}}}\big)\frac{f(E^+_{\mathbf{p}})-f(E^-_{\mathbf{p}})}{E^+_{\mathbf{p}}-E^-_{\mathbf{p}}}
\Big(\frac{1}{\omega-E^+_{\mathbf{p}}+E^-_{\mathbf{p}}}-\frac{1}{\omega+E^+_{\mathbf{p}}-E^-_{\mathbf{p}}}\Big)\Big\}.
\end{eqnarray} Hence \begin{eqnarray}\label{fsE2} &
&-\int_{-\infty}^{+\infty}d\omega\frac{1}{\pi}\textrm{Im}\big[\mathbf{q}\cdot\mathbf{Q}^0_{33}(\omega,\mathbf{q})\big]\nonumber\\
&=&\int_{-\infty}^{+\infty}d\omega\sum_{\mathbf{p}}\frac{\omega}{2}\frac{\mathbf{p}\cdot\mathbf{q}}{m}\Big\{\big(\frac{\xi^+_{\mathbf{p}}}{E^+_{\mathbf{p}}}-\frac{\xi^-_{\mathbf{p}}}{E^-_{\mathbf{p}}}\big)\frac{1-f(E^+_{\mathbf{p}})-f(E^-_{\mathbf{p}})}{E^+_{\mathbf{p}}+E^-_{\mathbf{p}}}
\Big(\delta(\omega-E^+_{\mathbf{p}}-E^-_{\mathbf{p}})-\delta(\omega+E^+_{\mathbf{p}}+E^-_{\mathbf{p}})\Big)\nonumber\\
&
&\qquad\qquad\qquad\qquad\quad-\big(\frac{\xi^+_{\mathbf{p}}}{E^+_{\mathbf{p}}}+\frac{\xi^-_{\mathbf{p}}}{E^-_{\mathbf{p}}}\big)\frac{f(E^+_{\mathbf{p}})-f(E^-_{\mathbf{p}})}{E^+_{\mathbf{p}}-E^-_{\mathbf{p}}}
\Big(\delta(\omega-E^+_{\mathbf{p}}+E^-_{\mathbf{p}})-\delta(\omega+E^+_{\mathbf{p}}-E^-_{\mathbf{p}})\Big)\Big\}\nonumber\\
&=&\sum_{\mathbf{p}}\frac{\mathbf{p}\cdot\mathbf{q}}{m}\big[1-\frac{\xi^-_{\mathbf{p}}}{E^-_{\mathbf{p}}}\big(1-2f(E^-_{\mathbf{p}})\big)\big]-2\sum_{\mathbf{p}}\frac{\mathbf{p}\cdot\mathbf{q}}{m}\big[1-\frac{\xi^+_{\mathbf{p}}}{E^+_{\mathbf{p}}}\big(1-2f(E^+_{\mathbf{p}})\big)\big].
\end{eqnarray}
We change variables by
$\mathbf{p}\rightarrow\mathbf{p}+\frac{\mathbf{q}}{2}$ in the first term, and change variables by $\mathbf{p}\rightarrow\mathbf{p}-\frac{\mathbf{q}}{2}$ in the
second term to get
 \begin{eqnarray}\label{fsE3}
-\int_{-\infty}^{+\infty}d\omega\frac{1}{\pi}\textrm{Im}\big[\mathbf{q}\cdot\mathbf{Q}^0_{33}(\omega,\mathbf{q})\big]
=\sum_{\mathbf{p}}\frac{\big[(\mathbf{p}+\frac{\mathbf{q}}{2})-(\mathbf{p}-\frac{\mathbf{q}}{2})\big]\cdot\mathbf{q}}{m}\Big[1-\frac{\xi_{\mathbf{p}}}{E_{\mathbf{p}}}\big(1-2f(E_{\mathbf{p}})\big)\Big]=\frac{q^2}{m}n,
 \end{eqnarray} where the number equation (\ref{NGE}) has been used.
This proves the lemma.

\section{Integral Equation of EM Vertex and GWI}\label{app:b0}
Here we prove that the vertex determined by the integral equation (\ref{IE}) obeys GWI (\ref{GWI5}). We will use the following equality
\begin{eqnarray}\label{tmp40}
g\sum_P\big(\sigma_3\hat{G}(P)-\hat{G}(P+Q)\sigma_3\big)=-(\sigma_3\hat{\Sigma}-\hat{\Sigma}\sigma_3).
\end{eqnarray}
To prove the proposition, we only need to show
\begin{eqnarray}
\sigma_3\hat{G}^{-1}(P+Q)-\hat{G}^{-1}(P)\sigma_3=q_{\mu}\hat{\gamma}^{\mu}(P+Q,P)+g\sum_K\sigma_3\hat{G}(K)q_{\mu}\hat{\Gamma}^{\mu}(K+Q,K)\hat{G}(K+Q)\sigma_3,
\end{eqnarray}
which is equivalent to
\begin{eqnarray}\label{dFD}
-(\sigma_3\hat{\Sigma}-\hat{\Sigma}\sigma_3)=g\sum_K\sigma_3\hat{G}(K)q_{\mu}\hat{\Gamma}^{\mu}(K+Q,K)\hat{G}(K+Q)\sigma_3.
\end{eqnarray}
From $\hat{\Sigma}=\hat{G}^{-1}_0(P)-\hat{G}^{-1}(P)$ one concludes that
\begin{eqnarray}\label{tmp21}
\hat{G}(P)\hat{G}^{-1}_0(P)=1+\hat{G}(P)\hat{\Sigma},\qquad\hat{G}^{-1}_0(P)\hat{G}(P)=1+\hat{\Sigma}\hat{G}(P).
\end{eqnarray}
Now we turn to the proof of Eq.(\ref{dFD}). By substituting Eq.(\ref{GWI5}) into its RHS and repeating the process, we get an iterative equation
\begin{eqnarray}\label{tmp41}
& &\mbox{RHS of Eq.~(\ref{dFD})}\nonumber\\
&=&g\sum_{P}\sigma_3\hat{G}(P)q_{\mu}\gamma^{\mu}(Q)\hat{G}(P+Q)\sigma_3\nonumber\\
& &+g^2\sum_{P_1P_2}\sigma_3\hat{G}(P_1)\sigma_3\hat{G}(P_2)q_{\mu}\gamma^{\mu}(P+Q,P)\hat{G}(P_2+Q)\sigma_3\hat{G}(P_1+Q)\sigma_3+\cdots\nonumber\\
&=&\sum_{i=1}^{\infty}g^{i}\sum_{P_1\cdots P_i}\sigma_3\hat{G}(P_1)\cdots\sigma_3\hat{G}(P_i)q_{\mu}\gamma^{\mu}(P_i+Q,P_i)\hat{G}(P_i+Q)\sigma_3\cdots\hat{G}(P_1+Q)\sigma_3\nonumber\\
&=&\sum_{i=1}^{\infty}g^{i}\sum_{P_1\cdots P_i}\prod_{k=1}^{i}\big[\sigma_3\hat{G}(P_k)\big]q_{\mu}\gamma^{\mu}(P_i+Q,P_i)\prod_{k=1}^{i}\big[\hat{G}(P_{i+1-k}+Q)\sigma_3\big],
\end{eqnarray}
After inserting the Ward identity (\ref{bGWI}) for the bare EM vertex and using Eqs.(\ref{tmp21}), we get
\begin{eqnarray}
& &\mbox{RHS of Eq.(\ref{dFD})}\nonumber\\
&=&\sum_{i=1}^{\infty}g^{i}\sum_{P_1\cdots P_i}\sigma_3\hat{G}(P_1)\cdots\sigma_3\hat{G}(P_i)\sigma_3\hat{G}_{0}^{-1}(P_i+Q)\hat{G}(P_i+Q)\sigma_3\cdots\hat{G}(P_1+Q)\sigma_3\nonumber\\
&-&\sum_{i=1}^{\infty}g^{i}\sum_{P_1\cdots P_i}\sigma_3\hat{G}(P_1)\cdots\sigma_3\hat{G}(P_i)\hat{G}_{0}^{-1}(P_i)\sigma_3\hat{G}(P_i+Q)\sigma_3\cdots\hat{G}(P_1+Q)\sigma_3\nonumber\\
&=&g\sum_{P}\big(\sigma_3\hat{G}(P)-\hat{G}(P+Q)\sigma_3\big)\nonumber\\
&+&\sum_{i=2}^{\infty}g^{i}\sum_{P_1\cdots P_i}\sigma_3\hat{G}(P_1)\cdots\sigma_3\hat{G}(P_i)\hat{G}(P_{i-1}+Q)\sigma_3\cdots\hat{G}(P_1+Q)\sigma_3\nonumber\\
&+&\sum_{i=1}^{\infty}g^{i}\sum_{P_1\cdots P_i}\sigma_3\hat{G}(P_1)\cdots\sigma_3\hat{G}(P_i)\sigma_3\hat{\Sigma}\hat{G}(P_{i}+Q)\sigma_3\cdots\hat{G}(P_1+Q)\sigma_3\nonumber\\
&-&\sum_{i=2}^{\infty}g^{i}\sum_{P_1\cdots P_i}\sigma_3\hat{G}(P_1)\cdots\sigma_3\hat{G}(P_{i-1})\hat{G}(P_{i}+Q)\sigma_3\cdots\hat{G}(P_1+Q)\sigma_3\nonumber\\
&-&\sum_{i=1}^{\infty}g^{i}\sum_{P_1\cdots P_i}\sigma_3\hat{G}(P_1)\cdots\sigma_3\hat{G}(P_i)\hat{\Sigma}\sigma_3\hat{G}(P_{i}+Q)\sigma_3\cdots\hat{G}(P_1+Q)\sigma_3.
\end{eqnarray}
By changing the dummy index $i\rightarrow i+1$ in the second and fourth summations, we get
\begin{eqnarray}
& &\mbox{RHS of Eq.(\ref{dFD})}\nonumber\\
&=&-(\sigma_3\hat{\Sigma}-\hat{\Sigma}\sigma_3)\nonumber\\
&+&\sum_{i=1}^{\infty}g^{i}\sum_{P_1\cdots P_i}\prod_{k=1}^{i}\big[\sigma_3\hat{G}(P_k)\big]g\sum_{P_{i+1}}\big[\sigma_3\hat{G}(P_{i+1})-\hat{G}(P_{i+1}+Q)\sigma_3\big]\prod_{k=1}^{i}\big[\hat{G}(P_{i+1-k}+Q)\sigma_3\big]\nonumber\\
&+&\sum_{i=1}^{\infty}g^{i}\sum_{P_1\cdots P_i}\prod_{k=1}^{i}\big[\sigma_3\hat{G}(P_k)\big]\big[\sigma_3\hat{\Sigma}-\hat{\Sigma}\sigma_3\big]\prod_{k=1}^{i}\big[\hat{G}(P_{i+1-k}+Q)\sigma_3\big]\nonumber\\
&=&-(\sigma_3\hat{\Sigma}-\hat{\Sigma}\sigma_3)=\textrm{LHS of Eq.(\ref{dFD})},
\end{eqnarray}
where Eq.(\ref{tmp40}) has been applied. Therefore we have proved that any vertex that satisfies the integral equation must also satisfy the Ward identity and hence must be gauge invariant.

\section{Evaluations of the Vertex}\label{app:b}
From the expressions given in Appendix~\ref{app:a0}, at $T=0$ we have
\begin{eqnarray}
& &\frac{2}{g}+Q_{22}(\omega,\mathbf{q})\simeq\frac{N(0)}{4}\int^{+\infty}_{-\infty}d\xi_{\mathbf{p}}\int^1_{-1}d\textrm{cos}\theta\frac{\omega^2-\frac{q^2p^2\textrm{cos}^2\theta}{m^2}}{E^2_{\mathbf{p}}}\frac{2E_{\mathbf{p}}}{-4E^2_{\mathbf{p}}}=-\frac{N(0)}{4}\frac{2}{\Delta^2}\big(\omega^2-\frac{2}{3}\frac{q^2\mu}{m}\big),
\end{eqnarray}
where in the fifth line we have used $p^2=2m(\xi_{\mathbf{p}}+\mu)$. Note that $\mu\simeq\epsilon_F=\frac{k^2_F}{2m}$. Thus
\begin{eqnarray}  \label{eqn:Q22}
& &\tilde{Q}_{22}(\omega,\mathbf{q})=-\frac{N(0)}{2\Delta^2}\big(\omega^2-\frac{1}{3}\frac{q^2k^2_F}{m^2}\big)=-\frac{N(0)}{2\Delta^2}\big(\omega^2-c^2_sq^2\big).
\end{eqnarray}
Similarly, for the temporal component of $Q^{\mu}_{23}$ we have
\begin{eqnarray}
Q_{23}(\omega,\mathbf{q})\simeq i\int\frac{d^3\mathbf{p}}{(2\pi)^3}\frac{\Delta\omega}{2E^2_{\mathbf{p}}}\frac{2E_{\mathbf{p}}}{-4E^2_{\mathbf{p}}}=-i\frac{N(0)\omega}{2\Delta},
\end{eqnarray}
Therefore $\Pi^0(\omega,\mathbf{q})=\frac{\Delta\omega}{\omega^2-c^2_sq^2}$ and the temporal component of the full vertex $\hat{\Gamma}^{\mu}$ is given by
\begin{eqnarray}
\hat{\Gamma}^0(P+Q,P)\simeq\sigma_3+2i\sigma_2\frac{\Delta\omega}{\omega^2-c^2_sq^2}.
\end{eqnarray}
For the spatial component, we have
\begin{eqnarray}
\mathbf{Q}_{23}(\omega,\mathbf{q})\simeq i\Delta\mathbf{q}\cdot\int\frac{d^3\mathbf{p}}{(2\pi)^3}\frac{\mathbf{p}\mathbf{p}}{m^2}\frac{1}{E^2_{\mathbf{p}}}\frac{2E_{\mathbf{p}}}{-4E^2_{\mathbf{p}}}=-\frac{i\Delta}{3m^2}\mathbf{q}\cdot\tensor{1}\int\frac{d^3\mathbf{p}}{(2\pi)^3}\frac{p^2}{2E^3_{\mathbf{p}}}\simeq-\frac{i\mathbf{q}}{3\Delta}\frac{2\mu N(0)}{m}.\nonumber\\
\end{eqnarray}
Using $\mu\simeq \frac{k^2_F}{2m}=\frac{3m}{2}c^2_s$, we have $\mathbf{Q}_{23}(\omega,\mathbf{q})\simeq-i\frac{N(0)c^2_s}{\Delta}\mathbf{q}$
, so
\begin{eqnarray}
\mathbf{\Pi}(\omega,\mathbf{q})\simeq\frac{2\Delta c^2_s}{\omega^2-c^2_sq^2}\mathbf{q}.
\end{eqnarray}
The spatial component is then given by
\begin{eqnarray}
\hat{\mathbf{\Gamma}}(P+Q,P)=\left(\begin{array}{cc}\frac{\mathbf{p}+\frac{\mathbf{q}}{2}}{m} & \mathbf{\Pi}(\omega,\mathbf{q}) \\ \bar{\mathbf{\Pi}}(\omega,\mathbf{q}) & \frac{\mathbf{p}+\frac{\mathbf{q}}{2}}{m}\end{array}\right)=\frac{\mathbf{p}+\frac{\mathbf{q}}{2}}{m}+2i\sigma_2\frac{\Delta c^2_s\mathbf{q}}{\omega^2-c^2_sq^2},
\end{eqnarray}

\section{Physical interpretation of the gauge-invariant linear response theory}\label{app:e}
In the main text we have seen that the WIs for
response functions are indeed satisfied even without knowing the exact form of the
full EM vertex. Here we pay some
attention to the gauge invariance of the CFOP linear response theory from the point of
view of a gauge transformation for the BCS Lagrangian. In real space, the Lagrangian
density following the BCS approximation is given by \begin{eqnarray}\label{tmp2}
\mathcal{L}_{\textrm{BCS}}=\Psi^{\dagger}\big(i\frac{\partial}{\partial
t}-(\frac{(-i\nabla)^2}{2m}-\mu)\sigma_3-A_{\mu}\hat{\gamma}^{\mu}+\Delta\sigma_1\big)\Psi,
\end{eqnarray} where $\hat{\gamma}^{\mu}=(\sigma_3,-\frac{i\nabla}{m})$. This Lagrangian density is obviously not invariant under the infinitesimal
gauge transformation $\Psi\rightarrow (1-i\sigma_3\chi)\Psi$,
$\Psi^{\dagger}\rightarrow \Psi^{\dagger}(1+i\sigma_3\chi)$ and $A_{\mu}\rightarrow
A_{\mu}+\partial_{\mu}\chi$ if the fluctuations of the order parameter are not
considered. Now we split the order parameter into its equilibrium and perturbative
parts as $\Delta\rightarrow\Delta+\Delta'$, where the equilibrium value is $\Delta$
and the perturbation is $\Delta'$. Therefore in the CFOP theory, the Lagrangian
density in real space becomes \begin{eqnarray}\label{LD}
\mathcal{L}_{\textrm{CFOP}}=\mathcal{L}_{\textrm{BCS}0}+\mathcal{L}'
=\Psi^{\dagger}\big(i\frac{\partial}
{\partial t}
-(\frac{(-i\nabla)^2}{2m}-\mu)\sigma_3+\Delta\sigma_1\big)\Psi-\Psi^{\dagger}\big
(\Delta
\sigma_1+\Delta_2\sigma_2+A_{\mu}\hat{\gamma}^{\mu}\big)\Psi,
\end{eqnarray}
Here the subscript ``0'' denotes the part in equilibrium. The gauge
transformation of $\Psi$ and $\Psi^{\dagger}$ leads to the fluctuations of the order
parameter $\delta\Delta_1=0$ and $\delta\Delta_2=-2\chi\Delta$. Therefore, the
following generalized infinitesimal gauge transformation leaves the Lagrangian
density (\ref{tmp2}) of the CFOP theory invariant:
\begin{eqnarray}\label{GT} &
&\Psi\rightarrow (1-i\sigma_3\chi)\Psi,\quad \Psi^{\dagger}\rightarrow
\Psi^{\dagger}(1+i\sigma_3\chi),\quad \Delta\rightarrow\Delta,\nonumber\\ &
&A_{\mu}\rightarrow A_{\mu}+\partial_{\mu}\chi,\quad
\Delta_1\rightarrow\Delta_1,\quad \Delta_2\rightarrow\Delta_2-2\Delta\chi.
\end{eqnarray} Under this generalized infinitesimal transformation the two parts of
the Lagrangian density transform according to \begin{eqnarray}
\mathcal{L}_{\textrm{BCS}0}&\rightarrow&
\mathcal{L}_{\textrm{BCS}0}+\Psi^{\dagger}\partial_{\mu}\chi\hat{\gamma}^{\mu}\Psi-i\chi\Psi^{
\dagger}\Delta
[\sigma_1,\sigma_3]\Psi
=\mathcal{L}_{BCS0}+\Psi^{\dagger}\partial_{\mu}\chi\hat{\gamma}^{\mu}\Psi-2\chi\Psi^{\dagger}
\Delta\sigma_2\Psi,
\nonumber\\ \mathcal{L}'&\rightarrow&\mathcal{L}'+2\chi\Psi^{\dagger}\Delta
\sigma_2\Psi-\Psi^{\dagger}\partial_{\mu}\chi\hat{\gamma}^{\mu}\Psi. \end{eqnarray}
Therefore $\mathcal{L}_{\textrm{BCS}}$ is indeed invariant under the generalized
infinitesimal gauge transformation (\ref{GT}). It is the gauge transformation of
$\Delta_2$ that compensates for the effects associated with the Cooper-pair condensation and
leads to the gauge invariance of the CFOP theory. The Noether current associated with
this generalized gauge transformation can be deduced by introducing the ``generalized
gauge space''. We define the space where the generalized external potential and
generalized interacting vertex (see Eq.~(\ref{GGP})) live as the generalized gauge space.
Explicitly, the perturbative Lagrangian density is rewritten as
-$\Psi^{\dagger}\hat{\mathbf{\Phi}}^T\cdot\hat{\mathbf{\Sigma}}\Psi$, where $\cdot$
denotes the inner product in this generalized gauge space. The generalized gauge
transformation of the generalized external potential is \begin{eqnarray} \label{GT2}
\hat{\mathbf{\Phi}}\rightarrow\hat{\mathbf{\Phi}}+\left(\begin{array}{c} 0 \\
-2\Delta\chi \\ \partial_{\mu}\chi\end{array}\right)\textrm{ or in the momentum space
}\hat{\mathbf{\Phi}}+\left(\begin{array}{c} 0 \\ -2\Delta\chi \\
-iq_{\mu}\chi\end{array}\right). \end{eqnarray} We define the generalized external
momentums $\hat{\mathbf{q}}\equiv(0,2i\Delta,q_{\mu})^T$ and
$\hat{\bar{\mathbf{q}}}\equiv(0,2i\Delta,-q_{\mu})^T$ in the generalized gauge space.
Then the generalized gauge transformation (\ref{GT2}) can be written as
\begin{eqnarray}\label{GT3}
\hat{\mathbf{\Phi}}\rightarrow\hat{\mathbf{\Phi}}+i\hat{\bar{\mathbf{q}}}\chi.
\end{eqnarray} The GWI (\ref{WI0}) can be expressed as
\begin{eqnarray}\label{GWI2}
\sigma_3\hat{G}^{-1}(P+Q)-\hat{G}^{-1}(P)\sigma_3=\hat{\mathbf{q}}^T\cdot{\hat{\mathbf{\Sigma}}}.
\end{eqnarray} In fact, this is the generalized Ward identity associated with the
generalized gauge transformation (\ref{GT2}) in the generalized gauge space.

Next we address the conserved current associated with this gauge transformation.
Using the self-consistent condition $\delta J_{1,2}=-\frac{2}{g}\Delta_{1,2}$, the
current in Eq.~(\ref{eqn:Q}) can be written as
\begin{eqnarray}\label{eqn:Q2} \left(
\begin{array}{c} 0 \\ 0 \\ J^{\mu} \end{array}\right)=\left( \begin{array}{ccc}
\tilde{Q}_{11} & Q_{12} & Q^{\nu}_{13} \\ Q_{21} & \tilde{Q}_{22} & Q^{\nu}_{23} \\
Q^{\mu}_{31} & Q^{\mu}_{32} & \tilde{Q}^{\mu\nu}_{33} \end{array}\right)\left(
\begin{array}{ccc} \Delta_{1} \\ \Delta_{2} \\A_{\nu} \end{array}\right).
\end{eqnarray} We then define the generalized current
$\hat{\mathbf{J}}\equiv(0,0,J^{\mu})^T$ and three generalized response-function vectors
\begin{eqnarray}\label{eqn:Q3} \hat{\mathbf{Q}}_1=\left( \begin{array}{c}
\tilde{Q}_{11} \\ Q_{21} \\ Q^{\mu}_{31} \end{array}\right),\qquad
\hat{\mathbf{Q}}_2=\left( \begin{array}{c} Q_{12} \\ \tilde{Q}_{22} \\ Q^{\mu}_{32}
\end{array}\right),\qquad \hat{\mathbf{Q}}^{\mu}_3=\left( \begin{array}{c}
Q^{\mu}_{13} \\ Q^{\mu}_{23} \\ \tilde{Q}^{\mu\nu}_{33} \end{array}\right).
\end{eqnarray} Then the current equation (\ref{eqn:Q2}) becomes
\begin{eqnarray}\label{CE0}
\hat{\mathbf{J}}=(\hat{\mathbf{Q}}_1,\hat{\mathbf{Q}}_2,\hat{\mathbf{Q}}^{\mu}_3)\cdot\hat{\mathbf{\Phi}},
\end{eqnarray} The GWIs (\ref{WI}) for the response functions can also be written as
\begin{eqnarray}\label{GWI3} \hat{\mathbf{q}}^T\cdot\hat{\mathbf{Q}}_i=0,
\quad\textrm{for $i=1,2,3$}. \end{eqnarray} Thus the GWIs directly lead to the
conservation of the generalized current \begin{eqnarray}\label{CE1}
\hat{\mathbf{q}}^T\cdot\hat{\mathbf{J}}=(\hat{\mathbf{q}}^T\cdot\hat{\mathbf{Q}}_1,\hat{\mathbf{q}}^T\cdot\hat{\mathbf{Q}}_2,\hat{\mathbf{q}}^T\cdot\hat{\mathbf{Q}}^{\mu}_3)\cdot\hat{\mathbf{\Phi}}=0.
\end{eqnarray} This gives the conservation law of the EM current $q_{\mu}J^{\mu}=0$.
Therefore $\hat{\mathbf{J}}$ is indeed the Neother current associated with the
generalized gauge transformation. Moreover, by noting that the GWIs (\ref{GWI3}) in
the generalized gauge space can be written as \begin{eqnarray}\label{CE3}
(\hat{\mathbf{Q}}_1,\hat{\mathbf{Q}}_2,\hat{\mathbf{Q}}^{\mu}_3)\cdot\hat{\bar{\mathbf{q}}}=0.
\end{eqnarray} Under the generalized gauge transformation (\ref{GT3}), the
generalized current transforms as \begin{eqnarray}\label{CE4}
\hat{\mathbf{J}}=(\hat{\mathbf{Q}}_1,\hat{\mathbf{Q}}_2,\hat{\mathbf{Q}}^{\mu}_3)\cdot\hat{\mathbf{\Phi}}\rightarrow
(\hat{\mathbf{Q}}_1,\hat{\mathbf{Q}}_2,\hat{\mathbf{Q}}^{\mu}_3)\cdot(\hat{\mathbf{\Phi}}+i\hat{\bar{\mathbf{q}}}\chi)=\hat{\mathbf{J}}.
\end{eqnarray} Thus the generalized current is indeed invariant under
the generalized gauge transformation.

\bibliographystyle{apsrev}

\end{document}